\pdfoutput=1
\documentclass[12pt,a4paper]{article}
\usepackage{ifthen} 
\newboolean{pdflatex}
\setboolean{pdflatex}{true} 
\newboolean{articletitles}
\setboolean{articletitles}{true} 
\newboolean{uprightparticles}
\setboolean{uprightparticles}{false} 

\def\paperauthors{T. Evans$^{1,2}$, C. Fitzpatrick$^1$, J. Horswill$^1$}
\def\addresses{$^1${Department of Physics and Astronomy, University of Manchester, Manchester, United Kingdom} \\ $^2${Nikhef National Institute for Subatomic Physics, Amsterdam, Netherlands}}

\def\papertitle{An Automated Bandwidth Division for the LHCb Upgrade Trigger} 

\usepackage[top=1in, bottom=1.25in, left=1in, right=1in]{geometry}

\usepackage{microtype}
\usepackage{lineno}  
\usepackage{xspace} 
\usepackage{caption} 

\usepackage{color}
\usepackage{colortbl}

\usepackage{amsmath} 
\usepackage{amssymb}
\usepackage{amsfonts}
\usepackage{upgreek} 

\newcommand*\patchAmsMathEnvironmentForLineno[1]{%
\expandafter\let\csname old#1\expandafter\endcsname\csname #1\endcsname
\expandafter\let\csname oldend#1\expandafter\endcsname\csname
end#1\endcsname
 \renewenvironment{#1}%
   {\linenomath\csname old#1\endcsname}%
   {\csname oldend#1\endcsname\endlinenomath}%
}
\newcommand*\patchBothAmsMathEnvironmentsForLineno[1]{%
  \patchAmsMathEnvironmentForLineno{#1}%
  \patchAmsMathEnvironmentForLineno{#1*}%
}
\AtBeginDocument{%
\patchBothAmsMathEnvironmentsForLineno{equation}%
\patchBothAmsMathEnvironmentsForLineno{align}%
\patchBothAmsMathEnvironmentsForLineno{flalign}%
\patchBothAmsMathEnvironmentsForLineno{alignat}%
\patchBothAmsMathEnvironmentsForLineno{gather}%
\patchBothAmsMathEnvironmentsForLineno{multline}%
\patchBothAmsMathEnvironmentsForLineno{eqnarray}%
}

\usepackage[colorinlistoftodos,textsize=scriptsize]{todonotes}

\usepackage[bottom,flushmargin,hang,multiple]{footmisc}

\usepackage{cite} 
\usepackage{mciteplus}

\usepackage{graphicx}
\usepackage{amsmath}
\usepackage{amsfonts}
\usepackage{amssymb}
\usepackage{hyperref}
\usepackage{hyperxmp}
\usepackage{url}
\usepackage{cite}
\usepackage{lineno}
\usepackage{subfig}
\usepackage{url,amsmath,booktabs,subfig,tikz}
\usepackage{array}
\usepackage{makecell}
\usepackage{changepage}
\usepackage{xspace} 
\usepackage{upgreek}

\begin{document}
\renewcommand{\thefootnote}{\fnsymbol{footnote}}
\setcounter{footnote}{1}

\begin{titlepage}
\pagenumbering{roman}

\vspace*{-1.5cm}
\vspace*{1.5cm}
\noindent

\vspace*{4.0cm}

{\normalfont\bfseries\boldmath\huge
\begin{center}
\papertitle 
\end{center}
}

\vspace*{2.0cm}

\begin{center}
\paperauthors

\vspace{2mm}
{ \it \footnotesize
\addresses
}

\vspace{4mm}

{ \it \footnotesize
ORCIDs: 0000-0003-3674-0812, 0000-0003-3016-1879, 0000-0002-9199-8616
}

\vspace{2mm}

{ \it \footnotesize
Contact Email: joshhorswill@gmail.com
}

\vspace{4mm}

\today
\end{center}

\vspace{\fill}
\begin{abstract}
  \noindent
The upgraded Large Hadron Collider beauty (LHCb) experiment is the first detector based at a hadron collider using a fully software based trigger.
The first `High Level Trigger' stage (HLT1) reduces the event rate from 30 MHz to approximately 1 MHz based on reconstruction criteria from the tracking system, and consists of $\mathcal{O}(100)$ trigger selections implemented on Graphics Processing Units (GPUs). 
These selections are further refined following the full offline-quality reconstruction at the second stage (HLT2) prior to saving for analysis.
An automated bandwidth division has been performed to equitably divide this 1 MHz HLT1 Output Rate (OR) between the signals of interest to the LHCb physics programme. 
This was achieved by optimising a set of trigger selections that maximise efficiency for signals of interest to LHCb while keeping the total HLT1 readout capped to a maximum.
The bandwidth division tool has been used to determine the optimal selection for 35 selection algorithms over 80 characteristic physics channels.
\end{abstract}

\vspace*{2.0cm}

\begin{center}
  Submitted to Comput Softw Big Sci.
\end{center}

\vspace{\fill}

\vspace*{2mm}

\end{titlepage}


\setcounter{page}{2}
\mbox{~}


\renewcommand{\thefootnote}{\arabic{footnote}}
\setcounter{footnote}{0}



\pagestyle{plain} 
\setcounter{page}{1}
\pagenumbering{arabic}

\def\lhcb   {\mbox{LHCb}\xspace}
\def\atlas  {\mbox{ATLAS}\xspace}
\def\cms    {\mbox{CMS}\xspace}
\def\alice  {\mbox{ALICE}\xspace}
\def\babar  {\mbox{BaBar}\xspace}
\def\belle  {\mbox{Belle}\xspace}
\def\belletwo {\mbox{Belle~II}\xspace}
\def\besiii {\mbox{BESIII}\xspace}
\def\cleo   {\mbox{CLEO}\xspace}
\def\cdf    {\mbox{CDF}\xspace}
\def\dzero  {\mbox{D0}\xspace}
\def\aleph  {\mbox{ALEPH}\xspace}
\def\delphi {\mbox{DELPHI}\xspace}
\def\opal   {\mbox{OPAL}\xspace}
\def\lthree {\mbox{L3}\xspace}
\def\sld    {\mbox{SLD}\xspace}
\def\cern {\mbox{CERN}\xspace}
\def\lhc    {\mbox{LHC}\xspace}
\def\lep    {\mbox{LEP}\xspace}
\def\tevatron {Tevatron\xspace}
\def\bfactories {\mbox{\B Factories}\xspace}
\def\bfactory   {\mbox{\B Factory}\xspace}
\def\upgradeone {\mbox{Upgrade~I}\xspace}
\def\upgradetwo {\mbox{Upgrade~II}\xspace}


\def\velo   {VELO\xspace}
\def\rich   {RICH\xspace}
\def\richone {RICH1\xspace}
\def\richtwo {RICH2\xspace}
\def\ttracker {TT\xspace}
\def\intr   {IT\xspace}
\def\st     {ST\xspace}
\def\ot     {OT\xspace}
\def\herschel {\mbox{\textsc{HeRSCheL}}\xspace}
\def\spd    {SPD\xspace}
\def\presh  {PS\xspace}
\def\ecal   {ECAL\xspace}
\def\hcal   {HCAL\xspace}
\def\MagUp {\mbox{\em Mag\kern -0.05em Up}\xspace}
\def\MagDown {\mbox{\em MagDown}\xspace}

\def\ode    {ODE\xspace}
\def\daq    {DAQ\xspace}
\def\tfc    {TFC\xspace}
\def\ecs    {ECS\xspace}
\def\lone   {L0\xspace}
\def\hlt    {HLT\xspace}
\def\hltone {HLT1\xspace}
\def\hlttwo {HLT2\xspace}


\ifthenelse{\boolean{uprightparticles}}%
{\def\Palpha      {\ensuremath{\upalpha}\xspace}
 \def\Pbeta       {\ensuremath{\upbeta}\xspace}
 \def\Pgamma      {\ensuremath{\upgamma}\xspace}                 
 \def\Pdelta      {\ensuremath{\updelta}\xspace}                 
 \def\Pepsilon    {\ensuremath{\upepsilon}\xspace}                 
 \def\Pvarepsilon {\ensuremath{\upvarepsilon}\xspace}                 
 \def\Pzeta       {\ensuremath{\upzeta}\xspace}                 
 \def\Peta        {\ensuremath{\upeta}\xspace}
 \def\Ptheta      {\ensuremath{\uptheta}\xspace}                 
 \def\Pvartheta   {\ensuremath{\upvartheta}\xspace}                 
 \def\Piota       {\ensuremath{\upiota}\xspace}                 
 \def\Pkappa      {\ensuremath{\upkappa}\xspace}                 
 \def\Plambda     {\ensuremath{\uplambda}\xspace}                 
 \def\Pmu         {\ensuremath{\upmu}\xspace}                 
 \def\Pnu         {\ensuremath{\upnu}\xspace}                 
 \def\Pxi         {\ensuremath{\upxi}\xspace}                 
 \def\Ppi         {\ensuremath{\uppi}\xspace}                 
 \def\Pvarpi      {\ensuremath{\upvarpi}\xspace}                 
 \def\Prho        {\ensuremath{\uprho}\xspace}                 
 \def\Pvarrho     {\ensuremath{\upvarrho}\xspace}                 
 \def\Ptau        {\ensuremath{\uptau}\xspace}                 
 \def\Pupsilon    {\ensuremath{\upupsilon}\xspace}                 
 \def\Pphi        {\ensuremath{\upphi}\xspace}                 
 \def\Pvarphi     {\ensuremath{\upvarphi}\xspace}                 
 \def\Pchi        {\ensuremath{\upchi}\xspace}                 
 \def\Ppsi        {\ensuremath{\uppsi}\xspace}                 
 \def\Pomega      {\ensuremath{\upomega}\xspace}                 

 \def\PDelta      {\ensuremath{\Delta}\xspace}                 
 \def\PXi         {\ensuremath{\Xi}\xspace}                 
 \def\PLambda     {\ensuremath{\Lambda}\xspace}                 
 \def\PSigma      {\ensuremath{\Sigma}\xspace}                 
 \def\POmega      {\ensuremath{\Omega}\xspace}                 
 \def\PUpsilon    {\ensuremath{\Upsilon}\xspace}
 \let\oldPi\Pi
 \def\PPi         {\ensuremath{\oldPi}\xspace}

 \def\PA      {\ensuremath{\mathrm{A}}\xspace}                 
 \def\PB      {\ensuremath{\mathrm{B}}\xspace}                 
 \def\PC      {\ensuremath{\mathrm{C}}\xspace}                 
 \def\PD      {\ensuremath{\mathrm{D}}\xspace}                 
 \def\PE      {\ensuremath{\mathrm{E}}\xspace}                 
 \def\PF      {\ensuremath{\mathrm{F}}\xspace}                 
 \def\PG      {\ensuremath{\mathrm{G}}\xspace}                 
 \def\PH      {\ensuremath{\mathrm{H}}\xspace}                 
 \def\PI      {\ensuremath{\mathrm{I}}\xspace}                 
 \def\PJ      {\ensuremath{\mathrm{J}}\xspace}                 
 \def\PK      {\ensuremath{\mathrm{K}}\xspace}                 
 \def\PL      {\ensuremath{\mathrm{L}}\xspace}                 
 \def\PM      {\ensuremath{\mathrm{M}}\xspace}                 
 \def\PN      {\ensuremath{\mathrm{N}}\xspace}                 
 \def\PO      {\ensuremath{\mathrm{O}}\xspace}                 
 \def\PP      {\ensuremath{\mathrm{P}}\xspace}                 
 \def\PQ      {\ensuremath{\mathrm{Q}}\xspace}                 
 \def\PR      {\ensuremath{\mathrm{R}}\xspace}                 
 \def\PS      {\ensuremath{\mathrm{S}}\xspace}                 
 \def\PT      {\ensuremath{\mathrm{T}}\xspace}                 
 \def\PU      {\ensuremath{\mathrm{U}}\xspace}                 
 \def\PV      {\ensuremath{\mathrm{V}}\xspace}                 
 \def\PW      {\ensuremath{\mathrm{W}}\xspace}                 
 \def\PX      {\ensuremath{\mathrm{X}}\xspace}                 
 \def\PY      {\ensuremath{\mathrm{Y}}\xspace}                 
 \def\PZ      {\ensuremath{\mathrm{Z}}\xspace}                 
 \def\Pa      {\ensuremath{\mathrm{a}}\xspace}                 
 \def\Pb      {\ensuremath{\mathrm{b}}\xspace}                 
 \def\Pc      {\ensuremath{\mathrm{c}}\xspace}                 
 \def\Pd      {\ensuremath{\mathrm{d}}\xspace}                 
 \def\Pe      {\ensuremath{\mathrm{e}}\xspace}                 
 \def\Pf      {\ensuremath{\mathrm{f}}\xspace}                 
 \def\Pg      {\ensuremath{\mathrm{g}}\xspace}                 
 \def\Ph      {\ensuremath{\mathrm{h}}\xspace}                 
 \def\Pi      {\ensuremath{\mathrm{i}}\xspace}                 
 \def\Pj      {\ensuremath{\mathrm{j}}\xspace}                 
 \def\Pk      {\ensuremath{\mathrm{k}}\xspace}                 
 \def\Pl      {\ensuremath{\mathrm{l}}\xspace}                 
 \def\Pm      {\ensuremath{\mathrm{m}}\xspace}                 
 \def\Pn      {\ensuremath{\mathrm{n}}\xspace}                 
 \def\Po      {\ensuremath{\mathrm{o}}\xspace}                 
 \def\Pp      {\ensuremath{\mathrm{p}}\xspace}                 
 \def\Pq      {\ensuremath{\mathrm{q}}\xspace}                 
 \def\Pr      {\ensuremath{\mathrm{r}}\xspace}                 
 \def\Ps      {\ensuremath{\mathrm{s}}\xspace}                 
 \def\Pt      {\ensuremath{\mathrm{t}}\xspace}                 
 \def\Pu      {\ensuremath{\mathrm{u}}\xspace}                 
 \def\Pv      {\ensuremath{\mathrm{v}}\xspace}                 
 \def\Pw      {\ensuremath{\mathrm{w}}\xspace}                 
 \def\Px      {\ensuremath{\mathrm{x}}\xspace}                 
 \def\Py      {\ensuremath{\mathrm{y}}\xspace}                 
 \def\Pz      {\ensuremath{\mathrm{z}}\xspace}                 
 \def\thebaroffset{0.0em}
}
{\def\Palpha      {\ensuremath{\alpha}\xspace}
 \def\Pbeta       {\ensuremath{\beta}\xspace}
 \def\Pgamma      {\ensuremath{\gamma}\xspace}                 
 \def\Pdelta      {\ensuremath{\delta}\xspace}                 
 \def\Pepsilon    {\ensuremath{\epsilon}\xspace}                 
 \def\Pvarepsilon {\ensuremath{\varepsilon}\xspace}                 
 \def\Pzeta       {\ensuremath{\zeta}\xspace}                 
 \def\Peta        {\ensuremath{\eta}\xspace}                 
 \def\Ptheta      {\ensuremath{\theta}\xspace}                 
 \def\Pvartheta   {\ensuremath{\vartheta}\xspace}                 
 \def\Piota       {\ensuremath{\iota}\xspace}                 
 \def\Pkappa      {\ensuremath{\kappa}\xspace}                 
 \def\Plambda     {\ensuremath{\lambda}\xspace}                 
 \def\Pmu         {\ensuremath{\mu}\xspace}                 
 \def\Pnu         {\ensuremath{\nu}\xspace}                 
 \def\Pxi         {\ensuremath{\xi}\xspace}                 
 \def\Ppi         {\ensuremath{\pi}\xspace}                 
 \def\Pvarpi      {\ensuremath{\varpi}\xspace}                 
 \def\Prho        {\ensuremath{\rho}\xspace}                 
 \def\Pvarrho     {\ensuremath{\varrho}\xspace}                 
 \def\Ptau        {\ensuremath{\tau}\xspace}                 
 \def\Pupsilon    {\ensuremath{\upsilon}\xspace}                 
 \def\Pphi        {\ensuremath{\phi}\xspace}                 
 \def\Pvarphi     {\ensuremath{\varphi}\xspace}                 
 \def\Pchi        {\ensuremath{\chi}\xspace}                 
 \def\Ppsi        {\ensuremath{\psi}\xspace}                 
 \def\Pomega      {\ensuremath{\omega}\xspace}                 
 \mathchardef\PDelta="7101
 \mathchardef\PXi="7104
 \mathchardef\PLambda="7103
 \mathchardef\PSigma="7106
 \mathchardef\POmega="710A
 \mathchardef\PUpsilon="7107
 \mathchardef\PPi="7105
 \def\PA      {\ensuremath{A}\xspace}                 
 \def\PB      {\ensuremath{B}\xspace}                 
 \def\PC      {\ensuremath{C}\xspace}                 
 \def\PD      {\ensuremath{D}\xspace}                 
 \def\PE      {\ensuremath{E}\xspace}                 
 \def\PF      {\ensuremath{F}\xspace}                 
 \def\PG      {\ensuremath{G}\xspace}                 
 \def\PH      {\ensuremath{H}\xspace}                 
 \def\PI      {\ensuremath{I}\xspace}                 
 \def\PJ      {\ensuremath{J}\xspace}                 
 \def\PK      {\ensuremath{K}\xspace}                 
 \def\PL      {\ensuremath{L}\xspace}                 
 \def\PM      {\ensuremath{M}\xspace}                 
 \def\PN      {\ensuremath{N}\xspace}                 
 \def\PO      {\ensuremath{O}\xspace}                 
 \def\PP      {\ensuremath{P}\xspace}                 
 \def\PQ      {\ensuremath{Q}\xspace}                 
 \def\PR      {\ensuremath{R}\xspace}                 
 \def\PS      {\ensuremath{S}\xspace}                 
 \def\PT      {\ensuremath{T}\xspace}                 
 \def\PU      {\ensuremath{U}\xspace}                 
 \def\PV      {\ensuremath{V}\xspace}                 
 \def\PW      {\ensuremath{W}\xspace}                 
 \def\PX      {\ensuremath{X}\xspace}                 
 \def\PY      {\ensuremath{Y}\xspace}                 
 \def\PZ      {\ensuremath{Z}\xspace}                 
 \def\Pa      {\ensuremath{a}\xspace}                 
 \def\Pb      {\ensuremath{b}\xspace}                 
 \def\Pc      {\ensuremath{c}\xspace}                 
 \def\Pd      {\ensuremath{d}\xspace}                 
 \def\Pe      {\ensuremath{e}\xspace}                 
 \def\Pf      {\ensuremath{f}\xspace}                 
 \def\Pg      {\ensuremath{g}\xspace}                 
 \def\Ph      {\ensuremath{h}\xspace}                 
 \def\Pi      {\ensuremath{i}\xspace}                 
 \def\Pj      {\ensuremath{j}\xspace}                 
 \def\Pk      {\ensuremath{k}\xspace}                 
 \def\Pl      {\ensuremath{l}\xspace}                 
 \def\Pm      {\ensuremath{m}\xspace}                 
 \def\Pn      {\ensuremath{n}\xspace}                 
 \def\Po      {\ensuremath{o}\xspace}                 
 \def\Pp      {\ensuremath{p}\xspace}                 
 \def\Pq      {\ensuremath{q}\xspace}                 
 \def\Pr      {\ensuremath{r}\xspace}                 
 \def\Ps      {\ensuremath{s}\xspace}                 
 \def\Pt      {\ensuremath{t}\xspace}                 
 \def\Pu      {\ensuremath{u}\xspace}                 
 \def\Pv      {\ensuremath{v}\xspace}                 
 \def\Pw      {\ensuremath{w}\xspace}                 
 \def\Px      {\ensuremath{x}\xspace}                 
 \def\Py      {\ensuremath{y}\xspace}                 
 \def\Pz      {\ensuremath{z}\xspace}
 \def\thebaroffset{0.18em}
}
\newcommand{\offsetoverline}[2][\thebaroffset]{\kern #1\overline{\kern -#1 #2}}%

\makeatletter
\ifcase \@ptsize \relax
  \newcommand{\miniscule}{\@setfontsize\miniscule{4}{5}}
\or
  \newcommand{\miniscule}{\@setfontsize\miniscule{5}{6}}
\or
  \newcommand{\miniscule}{\@setfontsize\miniscule{5}{6}}
\fi
\makeatother

\DeclareRobustCommand{\optbar}[1]{\shortstack{{\miniscule (\rule[.5ex]{1.25em}{.18mm})}
  \\ [-.7ex] $#1$}}


\let\emi\en
\def\electron   {{\ensuremath{\Pe}}\xspace}
\def\en         {{\ensuremath{\Pe^-}}\xspace}   
\def\ep         {{\ensuremath{\Pe^+}}\xspace}
\def\epm        {{\ensuremath{\Pe^\pm}}\xspace} 
\def\emp        {{\ensuremath{\Pe^\mp}}\xspace} 
\def\epem       {{\ensuremath{\Pe^+\Pe^-}}\xspace}

\def\muon       {{\ensuremath{\Pmu}}\xspace}
\def\mup        {{\ensuremath{\Pmu^+}}\xspace}
\def\mun        {{\ensuremath{\Pmu^-}}\xspace} 
\def\mupm       {{\ensuremath{\Pmu^\pm}}\xspace} 
\def\mump       {{\ensuremath{\Pmu^\mp}}\xspace} 
\def\mumu       {{\ensuremath{\Pmu^+\Pmu^-}}\xspace}

\def\tauon      {{\ensuremath{\Ptau}}\xspace}
\def\taup       {{\ensuremath{\Ptau^+}}\xspace}
\def\taum       {{\ensuremath{\Ptau^-}}\xspace}
\def\taupm      {{\ensuremath{\Ptau^\pm}}\xspace}
\def\taump      {{\ensuremath{\Ptau^\mp}}\xspace}
\def\tautau     {{\ensuremath{\Ptau^+\Ptau^-}}\xspace}

\def\lepton     {{\ensuremath{\ell}}\xspace}
\def\ellm       {{\ensuremath{\ell^-}}\xspace}
\def\ellp       {{\ensuremath{\ell^+}}\xspace}
\def\ellpm      {{\ensuremath{\ell^\pm}}\xspace}
\def\ellmp      {{\ensuremath{\ell^\mp}}\xspace}
\def\ellell     {\ensuremath{\ell^+ \ell^-}\xspace}

\def\neu        {{\ensuremath{\Pnu}}\xspace}
\def\neub       {{\ensuremath{\overline{\Pnu}}}\xspace}
\def\neue       {{\ensuremath{\neu_e}}\xspace}
\def\neueb      {{\ensuremath{\neub_e}}\xspace}
\def\neum       {{\ensuremath{\neu_\mu}}\xspace}
\def\neumb      {{\ensuremath{\neub_\mu}}\xspace}
\def\neut       {{\ensuremath{\neu_\tau}}\xspace}
\def\neutb      {{\ensuremath{\neub_\tau}}\xspace}
\def\neul       {{\ensuremath{\neu_\ell}}\xspace}
\def\neulb      {{\ensuremath{\neub_\ell}}\xspace}


\def\g      {{\ensuremath{\Pgamma}}\xspace}
\def\H      {{\ensuremath{\PH^0}}\xspace}
\def\Hp     {{\ensuremath{\PH^+}}\xspace}
\def\Hm     {{\ensuremath{\PH^-}}\xspace}
\def\Hpm    {{\ensuremath{\PH^\pm}}\xspace}
\def\W      {{\ensuremath{\PW}}\xspace}
\def\Wp     {{\ensuremath{\PW^+}}\xspace}
\def\Wm     {{\ensuremath{\PW^-}}\xspace}
\def\Wpm    {{\ensuremath{\PW^\pm}}\xspace}
\def\Z      {{\ensuremath{\PZ}}\xspace}


\def\quark     {{\ensuremath{\Pq}}\xspace}
\def\quarkbar  {{\ensuremath{\overline \quark}}\xspace}
\def\qqbar     {{\ensuremath{\quark\quarkbar}}\xspace}
\def\uquark    {{\ensuremath{\Pu}}\xspace}
\def\uquarkbar {{\ensuremath{\overline \uquark}}\xspace}
\def\uubar     {{\ensuremath{\uquark\uquarkbar}}\xspace}
\def\dquark    {{\ensuremath{\Pd}}\xspace}
\def\dquarkbar {{\ensuremath{\overline \dquark}}\xspace}
\def\ddbar     {{\ensuremath{\dquark\dquarkbar}}\xspace}
\def\squark    {{\ensuremath{\Ps}}\xspace}
\def\squarkbar {{\ensuremath{\overline \squark}}\xspace}
\def\ssbar     {{\ensuremath{\squark\squarkbar}}\xspace}
\def\cquark    {{\ensuremath{\Pc}}\xspace}
\def\cquarkbar {{\ensuremath{\overline \cquark}}\xspace}
\def\ccbar     {{\ensuremath{\cquark\cquarkbar}}\xspace}
\def\bquark    {{\ensuremath{\Pb}}\xspace}
\def\bquarkbar {{\ensuremath{\overline \bquark}}\xspace}
\def\bbbar     {{\ensuremath{\bquark\bquarkbar}}\xspace}
\def\tquark    {{\ensuremath{\Pt}}\xspace}
\def\tquarkbar {{\ensuremath{\overline \tquark}}\xspace}
\def\ttbar     {{\ensuremath{\tquark\tquarkbar}}\xspace}


\def\hadron {{\ensuremath{\Ph}}\xspace}
\def\pion   {{\ensuremath{\Ppi}}\xspace}
\def\piz    {{\ensuremath{\pion^0}}\xspace}
\def\pip    {{\ensuremath{\pion^+}}\xspace}
\def\pim    {{\ensuremath{\pion^-}}\xspace}
\def\pipm   {{\ensuremath{\pion^\pm}}\xspace}
\def\pimp   {{\ensuremath{\pion^\mp}}\xspace}

\def\rhomeson {{\ensuremath{\Prho}}\xspace}
\def\rhoz     {{\ensuremath{\rhomeson^0}}\xspace}
\def\rhop     {{\ensuremath{\rhomeson^+}}\xspace}
\def\rhom     {{\ensuremath{\rhomeson^-}}\xspace}
\def\rhopm    {{\ensuremath{\rhomeson^\pm}}\xspace}
\def\rhomp    {{\ensuremath{\rhomeson^\mp}}\xspace}

\def\kaon    {{\ensuremath{\PK}}\xspace}
\def\Kbar    {{\ensuremath{\offsetoverline{\PK}}}\xspace}
\def\Kb      {{\ensuremath{\Kbar}}\xspace}
\def\KorKbar {\kern \thebaroffset\optbar{\kern -\thebaroffset \PK}{}\xspace}
\def\Kz      {{\ensuremath{\kaon^0}}\xspace}
\def\Kzb     {{\ensuremath{\Kbar{}^0}}\xspace}
\def\Kp      {{\ensuremath{\kaon^+}}\xspace}
\def\Km      {{\ensuremath{\kaon^-}}\xspace}
\def\Kpm     {{\ensuremath{\kaon^\pm}}\xspace}
\def\Kmp     {{\ensuremath{\kaon^\mp}}\xspace}
\def\KS      {{\ensuremath{\kaon^0_{\mathrm{S}}}}\xspace}
\def\Vzero   {{\ensuremath{V^0}}\xspace}
\def\KL      {{\ensuremath{\kaon^0_{\mathrm{L}}}}\xspace}
\def\Kstarz  {{\ensuremath{\kaon^{*0}}}\xspace}
\def\Kstarzb {{\ensuremath{\Kbar{}^{*0}}}\xspace}
\def\Kstar   {{\ensuremath{\kaon^*}}\xspace}
\def\Kstarb  {{\ensuremath{\Kbar{}^*}}\xspace}
\def\Kstarp  {{\ensuremath{\kaon^{*+}}}\xspace}
\def\Kstarm  {{\ensuremath{\kaon^{*-}}}\xspace}
\def\Kstarpm {{\ensuremath{\kaon^{*\pm}}}\xspace}
\def\Kstarmp {{\ensuremath{\kaon^{*\mp}}}\xspace}
\def\KorKbarz {\ensuremath{\KorKbar^0}\xspace}

\newcommand{\etaz}{\ensuremath{\Peta}\xspace}
\newcommand{\etapr}{\ensuremath{\Peta^{\prime}}\xspace}
\newcommand{\phiz}{\ensuremath{\Pphi}\xspace}
\newcommand{\omegaz}{\ensuremath{\Pomega}\xspace}


\def\Dbar    {{\ensuremath{\offsetoverline{\PD}}}\xspace}
\def\D       {{\ensuremath{\PD}}\xspace}
\def\Db      {{\ensuremath{\Dbar}}\xspace}
\def\DorDbar {\kern \thebaroffset\optbar{\kern -\thebaroffset \PD}\xspace}
\def\Dz      {{\ensuremath{\D^0}}\xspace}
\def\Dzb     {{\ensuremath{\Dbar{}^0}}\xspace}
\def\Dp      {{\ensuremath{\D^+}}\xspace}
\def\Dm      {{\ensuremath{\D^-}}\xspace}
\def\Dpm     {{\ensuremath{\D^\pm}}\xspace}
\def\Dmp     {{\ensuremath{\D^\mp}}\xspace}
\def\DpDm    {\ensuremath{\Dp {\kern -0.16em \Dm}}\xspace}
\def\Dstar   {{\ensuremath{\D^*}}\xspace}
\def\Dstarb  {{\ensuremath{\Dbar{}^*}}\xspace}
\def\Dstarz  {{\ensuremath{\D^{*0}}}\xspace}
\def\Dstarzb {{\ensuremath{\Dbar{}^{*0}}}\xspace}
\def\theDstarz{{\ensuremath{\D^{*}(2007)^{0}}}\xspace}
\def\theDstarzb{{\ensuremath{\Dbar^{*}(2007)^{0}}}\xspace}
\def\Dstarp  {{\ensuremath{\D^{*+}}}\xspace}
\def\Dstarm  {{\ensuremath{\D^{*-}}}\xspace}
\def\Dstarpm {{\ensuremath{\D^{*\pm}}}\xspace}
\def\Dstarmp {{\ensuremath{\D^{*\mp}}}\xspace}
\def\theDstarp{{\ensuremath{\D^{*}(2010)^{+}}}\xspace}
\def\theDstarm{{\ensuremath{\D^{*}(2010)^{-}}}\xspace}
\def\theDstarpm{{\ensuremath{\D^{*}(2010)^{\pm}}}\xspace}
\def\theDstarmp{{\ensuremath{\D^{*}(2010)^{\mp}}}\xspace}
\def\Ds      {{\ensuremath{\D^+_\squark}}\xspace}
\def\Dsp     {{\ensuremath{\D^+_\squark}}\xspace}
\def\Dsm     {{\ensuremath{\D^-_\squark}}\xspace}
\def\Dspm    {{\ensuremath{\D^{\pm}_\squark}}\xspace}
\def\Dsmp    {{\ensuremath{\D^{\mp}_\squark}}\xspace}
\def\Dss     {{\ensuremath{\D^{*+}_\squark}}\xspace}
\def\Dssp    {{\ensuremath{\D^{*+}_\squark}}\xspace}
\def\Dssm    {{\ensuremath{\D^{*-}_\squark}}\xspace}
\def\Dsspm   {{\ensuremath{\D^{*\pm}_\squark}}\xspace}
\def\Dssmp   {{\ensuremath{\D^{*\mp}_\squark}}\xspace}
\def\DporDsp {{\ensuremath{\D_{(\squark)}^+}}\xspace}
\def\DmorDsm {{\ensuremath{\D{}_{(\squark)}^-}}\xspace}
\def\DpmorDspm {{\ensuremath{\D{}_{(\squark)}^\pm}}\xspace}

\def\B       {{\ensuremath{\PB}}\xspace}
\def\Bbar    {{\ensuremath{\offsetoverline{\PB}}}\xspace}
\def\Bb      {{\ensuremath{\Bbar}}\xspace}
\def\BorBbar {\kern \thebaroffset\optbar{\kern -\thebaroffset \PB}\xspace}
\def\Bz      {{\ensuremath{\B^0}}\xspace}
\def\Bzb     {{\ensuremath{\Bbar{}^0}}\xspace}
\def\Bd      {{\ensuremath{\B^0}}\xspace}
\def\Bdb     {{\ensuremath{\Bbar{}^0}}\xspace}
\def\BdorBdbar {\kern \thebaroffset\optbar{\kern -\thebaroffset \Bd}\xspace}
\def\Bu      {{\ensuremath{\B^+}}\xspace}
\def\Bub     {{\ensuremath{\B^-}}\xspace}
\def\Bp      {{\ensuremath{\Bu}}\xspace}
\def\Bm      {{\ensuremath{\Bub}}\xspace}
\def\Bpm     {{\ensuremath{\B^\pm}}\xspace}
\def\Bmp     {{\ensuremath{\B^\mp}}\xspace}
\def\Bs      {{\ensuremath{\B^0_\squark}}\xspace}
\def\Bsb     {{\ensuremath{\Bbar{}^0_\squark}}\xspace}
\def\BsorBsbar {\kern \thebaroffset\optbar{\kern -\thebaroffset \Bs}\xspace}
\def\Bc      {{\ensuremath{\B_\cquark^+}}\xspace}
\def\Bcp     {{\ensuremath{\B_\cquark^+}}\xspace}
\def\Bcm     {{\ensuremath{\B_\cquark^-}}\xspace}
\def\Bcpm    {{\ensuremath{\B_\cquark^\pm}}\xspace}
\def\Bds     {{\ensuremath{\B_{(\squark)}^0}}\xspace}
\def\Bdsb    {{\ensuremath{\Bbar{}_{(\squark)}^0}}\xspace}
\def\BdorBs  {\Bds}
\def\BdorBsbar  {\Bdsb}


\def\jpsi     {{\ensuremath{{\PJ\mskip -3mu/\mskip -2mu\Ppsi}}}\xspace}
\def\psitwos  {{\ensuremath{\Ppsi{(2S)}}}\xspace}
\def\psiprpr  {{\ensuremath{\Ppsi(3770)}}\xspace}
\def\etac     {{\ensuremath{\Peta_\cquark}}\xspace}
\def\psires  {{\ensuremath{\Ppsi}}\xspace}
\def\chic     {{\ensuremath{\Pchi_\cquark}}\xspace}
\def\chiczero {{\ensuremath{\Pchi_{\cquark 0}}}\xspace}
\def\chicone  {{\ensuremath{\Pchi_{\cquark 1}}}\xspace}
\def\chictwo  {{\ensuremath{\Pchi_{\cquark 2}}}\xspace}
\def\chicJ    {{\ensuremath{\Pchi_{\cquark J}}}\xspace}
\def\Upsilonres  {{\ensuremath{\PUpsilon}}\xspace}
\def\Y#1S{\ensuremath{\PUpsilon{(#1S)}}\xspace}
\def\OneS  {{\Y1S}\xspace}
\def\TwoS  {{\Y2S}\xspace}
\def\ThreeS{{\Y3S}\xspace}
\def\FourS {{\Y4S}\xspace}
\def\FiveS {{\Y5S}\xspace}
\def\chib     {{\ensuremath{\Pchi_{b}}}\xspace}
\def\chibzero {{\ensuremath{\Pchi_{\bquark 0}}}\xspace}
\def\chibone  {{\ensuremath{\Pchi_{\bquark 1}}}\xspace}
\def\chibtwo  {{\ensuremath{\Pchi_{\bquark 2}}}\xspace}
\def\chibJ    {{\ensuremath{\Pchi_{\bquark J}}}\xspace}
\def\theX     {{\ensuremath{\Pchi_{c1}(3872)}}\xspace}


\def\proton      {{\ensuremath{\Pp}}\xspace}
\def\antiproton  {{\ensuremath{\overline \proton}}\xspace}
\def\neutron     {{\ensuremath{\Pn}}\xspace}
\def\antineutron {{\ensuremath{\overline \neutron}}\xspace}
\def\Deltares    {{\ensuremath{\PDelta}}\xspace}
\def\Deltaresbar {{\ensuremath{\overline \Deltares}}\xspace}
\def\Lz          {{\ensuremath{\PLambda}}\xspace}
\def\Lbar        {{\ensuremath{\offsetoverline{\PLambda}}}\xspace}
\def\LorLbar     {\kern \thebaroffset\optbar{\kern -\thebaroffset \PLambda}\xspace}
\def\Lambdares   {{\ensuremath{\PLambda}}\xspace}
\def\Lambdaresbar{{\ensuremath{\Lbar}}\xspace}
\def\Sigmares    {{\ensuremath{\PSigma}}\xspace}
\def\Sigmaz      {{\ensuremath{\Sigmares{}^0}}\xspace}
\def\Sigmap      {{\ensuremath{\Sigmares{}^+}}\xspace}
\def\Sigmam      {{\ensuremath{\Sigmares{}^-}}\xspace}
\def\Sigmaresbar {{\ensuremath{\offsetoverline{\Sigmares}}}\xspace}
\def\Sigmabarz   {{\ensuremath{\Sigmaresbar{}^0}}\xspace}
\def\Sigmabarp   {{\ensuremath{\Sigmaresbar{}^+}}\xspace}
\def\Sigmabarm   {{\ensuremath{\Sigmaresbar{}^-}}\xspace}
\def\Xires       {{\ensuremath{\PXi}}\xspace}
\def\Xiz         {{\ensuremath{\Xires^0}}\xspace}
\def\Xim         {{\ensuremath{\Xires^-}}\xspace}
\def\Xiresbar       {{\ensuremath{\offsetoverline{\Xires}}}\xspace}
\def\Xibarz      {{\ensuremath{\Xiresbar^0}}\xspace}
\def\Xibarp      {{\ensuremath{\Xiresbar^+}}\xspace}
\def\Omegares    {{\ensuremath{\POmega}}\xspace}
\def\Omegaresbar {{\ensuremath{\offsetoverline{\POmega}}}\xspace}
\def\Omegam      {{\ensuremath{\Omegares^-}}\xspace}
\def\Omegabarp   {{\ensuremath{\Omegaresbar^+}}\xspace}

\def\Lc          {{\ensuremath{\Lz^+_\cquark}}\xspace}
\def\Lcbar       {{\ensuremath{\Lbar{}^-_\cquark}}\xspace}
\def\Xic         {{\ensuremath{\Xires_\cquark}}\xspace}
\def\Xicz        {{\ensuremath{\Xires^0_\cquark}}\xspace}
\def\Xicp        {{\ensuremath{\Xires^+_\cquark}}\xspace}
\def\Xicbar      {{\ensuremath{\Xiresbar{}_\cquark}}\xspace}
\def\Xicbarz     {{\ensuremath{\Xiresbar{}_\cquark^0}}\xspace}
\def\Xicbarm     {{\ensuremath{\Xiresbar{}_\cquark^-}}\xspace}
\def\Omegac      {{\ensuremath{\Omegares^0_\cquark}}\xspace}
\def\Omegacbar   {{\ensuremath{\Omegaresbar{}_\cquark^0}}\xspace}
\def\Xicc        {{\ensuremath{\Xires_{\cquark\cquark}}}\xspace}
\def\Xiccbar     {{\ensuremath{\Xiresbar{}_{\cquark\cquark}}}\xspace}
\def\Xiccp       {{\ensuremath{\Xires^+_{\cquark\cquark}}}\xspace}
\def\Xiccpp      {{\ensuremath{\Xires^{++}_{\cquark\cquark}}}\xspace}
\def\Xiccbarm    {{\ensuremath{\Xiresbar{}_{\cquark\cquark}^-}}\xspace}
\def\Xiccbarmm   {{\ensuremath{\Xiresbar{}_{\cquark\cquark}^{--}}}\xspace}
\def\Omegacc     {{\ensuremath{\Omegares^+_{\cquark\cquark}}}\xspace}
\def\Omegaccbar  {{\ensuremath{\Omegaresbar{}_{\cquark\cquark}^-}}\xspace}
\def\Omegaccc    {{\ensuremath{\Omegares^{++}_{\cquark\cquark\cquark}}}\xspace}
\def\Omegacccbar {{\ensuremath{\Omegaresbar{}_{\cquark\cquark\cquark}^{--}}}\xspace}

\def\Lb           {{\ensuremath{\Lz^0_\bquark}}\xspace}
\def\Lbbar        {{\ensuremath{\Lbar{}^0_\bquark}}\xspace}
\def\Sigmab       {{\ensuremath{\Sigmares_\bquark}}\xspace}
\def\Sigmabp      {{\ensuremath{\Sigmares_\bquark^+}}\xspace}
\def\Sigmabz      {{\ensuremath{\Sigmares_\bquark^0}}\xspace}
\def\Sigmabm      {{\ensuremath{\Sigmares_\bquark^-}}\xspace}
\def\Sigmabpm     {{\ensuremath{\Sigmares_\bquark^\pm}}\xspace}
\def\Sigmabbar    {{\ensuremath{\Sigmaresbar_\bquark}}\xspace}
\def\Sigmabbarp   {{\ensuremath{\Sigmaresbar_\bquark^+}}\xspace}
\def\Sigmabbarz   {{\ensuremath{\Sigmaresbar_\bquark^0}}\xspace}
\def\Sigmabbarm   {{\ensuremath{\Sigmaresbar_\bquark^-}}\xspace}
\def\Sigmabbarpm  {{\ensuremath{\Sigmaresbar_\bquark^-}}\xspace}
\def\Xib          {{\ensuremath{\Xires_\bquark}}\xspace}
\def\Xibz         {{\ensuremath{\Xires^0_\bquark}}\xspace}
\def\Xibm         {{\ensuremath{\Xires^-_\bquark}}\xspace}
\def\Xibbar       {{\ensuremath{\Xiresbar{}_\bquark}}\xspace}
\def\Xibbarz      {{\ensuremath{\Xiresbar{}_\bquark^0}}\xspace}
\def\Xibbarp      {{\ensuremath{\Xiresbar{}_\bquark^+}}\xspace}
\def\Omegab       {{\ensuremath{\Omegares^-_\bquark}}\xspace}
\def\Omegabbar    {{\ensuremath{\Omegaresbar{}_\bquark^+}}\xspace}


\def\BF         {{\ensuremath{\mathcal{B}}}\xspace}
\def\BR         {\BF}
\def\BRvis      {{\ensuremath{\BR_{\mathrm{{vis}}}}}}
\newcommand{\decay}[2]{\ensuremath{#1\!\to #2}\xspace} 
\def\ra                 {\ensuremath{\rightarrow}\xspace}
\def\to                 {\ensuremath{\rightarrow}\xspace}

\newcommand{\tauBs}{{\ensuremath{\tau_{\Bs}}}\xspace}
\newcommand{\tauBd}{{\ensuremath{\tau_{\Bd}}}\xspace}
\newcommand{\tauBz}{{\ensuremath{\tau_{\Bz}}}\xspace}
\newcommand{\tauBu}{{\ensuremath{\tau_{\Bp}}}\xspace}
\newcommand{\tauDp}{{\ensuremath{\tau_{\Dp}}}\xspace}
\newcommand{\tauDz}{{\ensuremath{\tau_{\Dz}}}\xspace}
\newcommand{\tauL}{{\ensuremath{\tau_{\mathrm{ L}}}}\xspace}
\newcommand{\tauH}{{\ensuremath{\tau_{\mathrm{ H}}}}\xspace}

\newcommand{\mBd}{{\ensuremath{m_{\Bd}}}\xspace}
\newcommand{\mBp}{{\ensuremath{m_{\Bp}}}\xspace}
\newcommand{\mBs}{{\ensuremath{m_{\Bs}}}\xspace}
\newcommand{\mBc}{{\ensuremath{m_{\Bc}}}\xspace}
\newcommand{\mLb}{{\ensuremath{m_{\Lb}}}\xspace}

\def\grpsuthree {{\ensuremath{\mathrm{SU}(3)}}\xspace}
\def\grpsutw    {{\ensuremath{\mathrm{SU}(2)}}\xspace}
\def\grpuone    {{\ensuremath{\mathrm{U}(1)}}\xspace}

\def\ssqtw   {{\ensuremath{\sin^{2}\!\theta_{\mathrm{W}}}}\xspace}
\def\csqtw   {{\ensuremath{\cos^{2}\!\theta_{\mathrm{W}}}}\xspace}
\def\stw     {{\ensuremath{\sin\theta_{\mathrm{W}}}}\xspace}
\def\ctw     {{\ensuremath{\cos\theta_{\mathrm{W}}}}\xspace}
\def\ssqtwef {{\ensuremath{{\sin}^{2}\theta_{\mathrm{W}}^{\mathrm{eff}}}}\xspace}
\def\csqtwef {{\ensuremath{{\cos}^{2}\theta_{\mathrm{W}}^{\mathrm{eff}}}}\xspace}
\def\stwef   {{\ensuremath{\sin\theta_{\mathrm{W}}^{\mathrm{eff}}}}\xspace}
\def\ctwef   {{\ensuremath{\cos\theta_{\mathrm{W}}^{\mathrm{eff}}}}\xspace}
\def\gv      {{\ensuremath{g_{\mbox{\tiny V}}}}\xspace}
\def\ga      {{\ensuremath{g_{\mbox{\tiny A}}}}\xspace}

\def\order   {{\ensuremath{\mathcal{O}}}\xspace}
\def\ordalph {{\ensuremath{\mathcal{O}(\alpha)}}\xspace}
\def\ordalsq {{\ensuremath{\mathcal{O}(\alpha^{2})}}\xspace}
\def\ordalcb {{\ensuremath{\mathcal{O}(\alpha^{3})}}\xspace}

\newcommand{\MSb}{{\ensuremath{\overline{\mathrm{MS}}}}\xspace}
\newcommand{\lqcd}{{\ensuremath{\Lambda_{\mathrm{QCD}}}}\xspace}
\def\qsq       {{\ensuremath{q^2}}\xspace}


\def\eps   {{\ensuremath{\varepsilon}}\xspace}
\def\epsK  {{\ensuremath{\varepsilon_K}}\xspace}
\def\epsB  {{\ensuremath{\varepsilon_B}}\xspace}
\def\epsp  {{\ensuremath{\varepsilon^\prime_K}}\xspace}

\def\CP                {{\ensuremath{C\!P}}\xspace}
\def\CPT               {{\ensuremath{C\!PT}}\xspace}
\def\T                 {{\ensuremath{T}}\xspace}

\def\rhobar {{\ensuremath{\overline \rho}}\xspace}
\def\etabar {{\ensuremath{\overline \eta}}\xspace}

\def\Vud  {{\ensuremath{V_{\uquark\dquark}^{\phantom{\ast}}}}\xspace}
\def\Vcd  {{\ensuremath{V_{\cquark\dquark}^{\phantom{\ast}}}}\xspace}
\def\Vtd  {{\ensuremath{V_{\tquark\dquark}^{\phantom{\ast}}}}\xspace}
\def\Vus  {{\ensuremath{V_{\uquark\squark}^{\phantom{\ast}}}}\xspace}
\def\Vcs  {{\ensuremath{V_{\cquark\squark}^{\phantom{\ast}}}}\xspace}
\def\Vts  {{\ensuremath{V_{\tquark\squark}^{\phantom{\ast}}}}\xspace}
\def\Vub  {{\ensuremath{V_{\uquark\bquark}^{\phantom{\ast}}}}\xspace}
\def\Vcb  {{\ensuremath{V_{\cquark\bquark}^{\phantom{\ast}}}}\xspace}
\def\Vtb  {{\ensuremath{V_{\tquark\bquark}^{\phantom{\ast}}}}\xspace}
\def\Vuds  {{\ensuremath{V_{\uquark\dquark}^\ast}}\xspace}
\def\Vcds  {{\ensuremath{V_{\cquark\dquark}^\ast}}\xspace}
\def\Vtds  {{\ensuremath{V_{\tquark\dquark}^\ast}}\xspace}
\def\Vuss  {{\ensuremath{V_{\uquark\squark}^\ast}}\xspace}
\def\Vcss  {{\ensuremath{V_{\cquark\squark}^\ast}}\xspace}
\def\Vtss  {{\ensuremath{V_{\tquark\squark}^\ast}}\xspace}
\def\Vubs  {{\ensuremath{V_{\uquark\bquark}^\ast}}\xspace}
\def\Vcbs  {{\ensuremath{V_{\cquark\bquark}^\ast}}\xspace}
\def\Vtbs  {{\ensuremath{V_{\tquark\bquark}^\ast}}\xspace}


\newcommand{\dms}{{\ensuremath{\Delta m_{\squark}}}\xspace}
\newcommand{\dmd}{{\ensuremath{\Delta m_{\dquark}}}\xspace}
\newcommand{\DG}{{\ensuremath{\Delta\Gamma}}\xspace}
\newcommand{\DGs}{{\ensuremath{\Delta\Gamma_{\squark}}}\xspace}
\newcommand{\DGd}{{\ensuremath{\Delta\Gamma_{\dquark}}}\xspace}
\newcommand{\Gs}{{\ensuremath{\Gamma_{\squark}}}\xspace}
\newcommand{\Gd}{{\ensuremath{\Gamma_{\dquark}}}\xspace}
\newcommand{\MBq}{{\ensuremath{M_{\B_\quark}}}\xspace}
\newcommand{\DGq}{{\ensuremath{\Delta\Gamma_{\quark}}}\xspace}
\newcommand{\Gq}{{\ensuremath{\Gamma_{\quark}}}\xspace}
\newcommand{\dmq}{{\ensuremath{\Delta m_{\quark}}}\xspace}
\newcommand{\GL}{{\ensuremath{\Gamma_{\mathrm{ L}}}}\xspace}
\newcommand{\GH}{{\ensuremath{\Gamma_{\mathrm{ H}}}}\xspace}
\newcommand{\DGsGs}{{\ensuremath{\Delta\Gamma_{\squark}/\Gamma_{\squark}}}\xspace}
\newcommand{\Delm}{{\mbox{$\Delta m $}}\xspace}
\newcommand{\ACP}{{\ensuremath{{\mathcal{A}}^{\CP}}}\xspace}
\newcommand{\Adir}{{\ensuremath{{\mathcal{A}}^{\mathrm{ dir}}}}\xspace}
\newcommand{\Amix}{{\ensuremath{{\mathcal{A}}^{\mathrm{ mix}}}}\xspace}
\newcommand{\ADelta}{{\ensuremath{{\mathcal{A}}^\Delta}}\xspace}
\newcommand{\phid}{{\ensuremath{\phi_{\dquark}}}\xspace}
\newcommand{\sinphid}{{\ensuremath{\sin\!\phid}}\xspace}
\newcommand{\phis}{{\ensuremath{\phi_{\squark}}}\xspace}
\newcommand{\betas}{{\ensuremath{\beta_{\squark}}}\xspace}
\newcommand{\sbetas}{{\ensuremath{\sigma(\beta_{\squark})}}\xspace}
\newcommand{\stbetas}{{\ensuremath{\sigma(2\beta_{\squark})}}\xspace}
\newcommand{\stphis}{{\ensuremath{\sigma(\phi_{\squark})}}\xspace}
\newcommand{\sinphis}{{\ensuremath{\sin\!\phis}}\xspace}

\newcommand{\edet}{{\ensuremath{\varepsilon_{\mathrm{ det}}}}\xspace}
\newcommand{\erec}{{\ensuremath{\varepsilon_{\mathrm{ rec/det}}}}\xspace}
\newcommand{\esel}{{\ensuremath{\varepsilon_{\mathrm{ sel/rec}}}}\xspace}
\newcommand{\etrg}{{\ensuremath{\varepsilon_{\mathrm{ trg/sel}}}}\xspace}
\newcommand{\etot}{{\ensuremath{\varepsilon_{\mathrm{ tot}}}}\xspace}

\newcommand{\mistag}{\ensuremath{\omega}\xspace}
\newcommand{\wcomb}{\ensuremath{\omega^{\mathrm{comb}}}\xspace}
\newcommand{\etag}{{\ensuremath{\varepsilon_{\mathrm{tag}}}}\xspace}
\newcommand{\etagcomb}{{\ensuremath{\varepsilon_{\mathrm{tag}}^{\mathrm{comb}}}}\xspace}
\newcommand{\effeff}{\ensuremath{\varepsilon_{\mathrm{eff}}}\xspace}
\newcommand{\effeffcomb}{\ensuremath{\varepsilon_{\mathrm{eff}}^{\mathrm{comb}}}\xspace}
\newcommand{\efftag}{{\ensuremath{\etag(1-2\omega)^2}}\xspace}
\newcommand{\effD}{{\ensuremath{\etag D^2}}\xspace}

\newcommand{\etagprompt}{{\ensuremath{\varepsilon_{\mathrm{ tag}}^{\mathrm{Pr}}}}\xspace}
\newcommand{\etagLL}{{\ensuremath{\varepsilon_{\mathrm{ tag}}^{\mathrm{LL}}}}\xspace}


\def\BdToKstmm    {\decay{\Bd}{\Kstarz\mup\mun}}
\def\BdbToKstmm   {\decay{\Bdb}{\Kstarzb\mup\mun}}

\def\BsToJPsiPhi  {\decay{\Bs}{\jpsi\phi}}
\def\BdToJPsiKst  {\decay{\Bd}{\jpsi\Kstarz}}
\def\BdbToJPsiKst {\decay{\Bdb}{\jpsi\Kstarzb}}

\def\BsPhiGam     {\decay{\Bs}{\phi \g}}
\def\BdKstGam     {\decay{\Bd}{\Kstarz \g}}

\def\BTohh        {\decay{\B}{\Ph^+ \Ph'^-}}
\def\BdTopipi     {\decay{\Bd}{\pip\pim}}
\def\BdToKpi      {\decay{\Bd}{\Kp\pim}}
\def\BsToKK       {\decay{\Bs}{\Kp\Km}}
\def\BsTopiK      {\decay{\Bs}{\pip\Km}}
\def\Cpipi        {\ensuremath{C_{\pip\pim}}\xspace}
\def\Spipi        {\ensuremath{S_{\pip\pim}}\xspace}
\def\CKK          {\ensuremath{C_{\Kp\Km}}\xspace}
\def\SKK          {\ensuremath{S_{\Kp\Km}}\xspace}
\def\ADGKK        {\ensuremath{A^{\DG}_{\Kp\Km}}\xspace}

\def\BdKstee  {\decay{\Bd}{\Kstarz\epem}}
\def\BdbKstee {\decay{\Bdb}{\Kstarzb\epem}}
\def\bsll     {\decay{\bquark}{\squark \ell^+ \ell^-}}
\def\AFB      {\ensuremath{A_{\mathrm{FB}}}\xspace}
\def\FL       {\ensuremath{F_{\mathrm{L}}}\xspace}
\def\AT#1     {\ensuremath{A_{\mathrm{T}}^{#1}}\xspace}           
\def\btosgam  {\decay{\bquark}{\squark \g}}
\def\btodgam  {\decay{\bquark}{\dquark \g}}
\def\Bsmm     {\decay{\Bs}{\mup\mun}}
\def\Bdmm     {\decay{\Bd}{\mup\mun}}
\def\Bsee     {\decay{\Bs}{\epem}}
\def\Bdee     {\decay{\Bd}{\epem}}
\def\ctl       {\ensuremath{\cos{\theta_\ell}}\xspace}
\def\ctk       {\ensuremath{\cos{\theta_K}}\xspace}

\def\C#1      {\ensuremath{\mathcal{C}_{#1}}\xspace}                       
\def\Cp#1     {\ensuremath{\mathcal{C}_{#1}^{'}}\xspace}                    
\def\Ceff#1   {\ensuremath{\mathcal{C}_{#1}^{\mathrm{(eff)}}}\xspace}        
\def\Cpeff#1  {\ensuremath{\mathcal{C}_{#1}^{'\mathrm{(eff)}}}\xspace}       
\def\Ope#1    {\ensuremath{\mathcal{O}_{#1}}\xspace}                       
\def\Opep#1   {\ensuremath{\mathcal{O}_{#1}^{'}}\xspace}                    


\def\xprime     {\ensuremath{x^{\prime}}\xspace}
\def\yprime     {\ensuremath{y^{\prime}}\xspace}
\def\ycp        {\ensuremath{y_{\CP}}\xspace}
\def\agamma     {\ensuremath{A_{\Gamma}}\xspace}
\def\dkpicf     {\decay{\Dz}{\Km\pip}}


\newcommand{\nospaceunit}[1]{\ensuremath{\text{#1}}}       
\newcommand{\aunit}[1]{\ensuremath{\text{\,#1}}}       

\newcommand{\tev}{\aunit{Te\kern -0.1em V}\xspace}
\newcommand{\gev}{\aunit{Ge\kern -0.1em V}\xspace}
\newcommand{\mev}{\aunit{Me\kern -0.1em V}\xspace}
\newcommand{\kev}{\aunit{ke\kern -0.1em V}\xspace}
\newcommand{\ev}{\aunit{e\kern -0.1em V}\xspace}
\newcommand{\gevgev}{\ensuremath{\gev^2}\xspace} 
\newcommand{\mevc}{\ensuremath{\aunit{Me\kern -0.1em V\!/}c}\xspace}
\newcommand{\gevc}{\ensuremath{\aunit{Ge\kern -0.1em V\!/}c}\xspace}
\newcommand{\mevcc}{\ensuremath{\aunit{Me\kern -0.1em V\!/}c^2}\xspace}
\newcommand{\gevcc}{\ensuremath{\aunit{Ge\kern -0.1em V\!/}c^2}\xspace}
\newcommand{\gevgevcc}{\ensuremath{\gev^2\!/c^2}\xspace} 
\newcommand{\gevgevcccc}{\ensuremath{\gev^2\!/c^4}\xspace} 

\def\km   {\aunit{km}\xspace}
\def\m    {\aunit{m}\xspace}
\def\ma   {\ensuremath{\aunit{m}^2}\xspace}
\def\cm   {\aunit{cm}\xspace}
\def\cma  {\ensuremath{\aunit{cm}^2}\xspace}
\def\mm   {\aunit{mm}\xspace}
\def\mma  {\ensuremath{\aunit{mm}^2}\xspace}
\def\mum  {\ensuremath{\,\upmu\nospaceunit{m}}\xspace}
\def\muma {\ensuremath{\,\upmu\nospaceunit{m}^2}\xspace}
\def\nm   {\aunit{nm}\xspace}
\def\fm   {\aunit{fm}\xspace}
\def\barn{\aunit{b}\xspace}
\def\mbarn{\aunit{mb}\xspace}
\def\mub{\ensuremath{\,\upmu\nospaceunit{b}}\xspace}
\def\nb {\aunit{nb}\xspace}
\def\invnb {\ensuremath{\nb^{-1}}\xspace}
\def\pb {\aunit{pb}\xspace}
\def\invpb {\ensuremath{\pb^{-1}}\xspace}
\def\fb   {\ensuremath{\aunit{fb}}\xspace}
\def\invfb   {\ensuremath{\fb^{-1}}\xspace}
\def\ab   {\ensuremath{\aunit{ab}}\xspace}
\def\invab   {\ensuremath{\ab^{-1}}\xspace}

\def\sec  {\ensuremath{\aunit{s}}\xspace}
\def\ms   {\ensuremath{\aunit{ms}}\xspace}
\def\mus  {\ensuremath{\,\upmu\nospaceunit{s}}\xspace}
\def\ns   {\ensuremath{\aunit{ns}}\xspace}
\def\ps   {\ensuremath{\aunit{ps}}\xspace}
\def\fs   {\aunit{fs}}

\def\mhz  {\ensuremath{\aunit{MHz}}\xspace}
\def\khz  {\ensuremath{\aunit{kHz}}\xspace}
\def\hz   {\ensuremath{\aunit{Hz}}\xspace}

\def\invps{\ensuremath{\ps^{-1}}\xspace}
\def\invns{\ensuremath{\ns^{-1}}\xspace}

\def\yr   {\aunit{yr}\xspace}
\def\hr   {\aunit{hr}\xspace}

\def\degc {\ensuremath{^\circ}{\text{C}}\xspace}
\def\degk {\aunit{K}\xspace}

\def\Xrad {\ensuremath{X_0}\xspace}
\def\NIL{\ensuremath{\lambda_{\rm int}}\xspace}
\def\mip {MIP\xspace}
\def\neutroneq {\ensuremath{n_\nospaceunit{eq}}\xspace}
\def\neqcmcm {\ensuremath{\neutroneq/\nospaceunit{cm}^2}\xspace}
\def\kRad {\aunit{kRad}\xspace}
\def\MRad {\aunit{MRad}\xspace}
\def\ci {\aunit{Ci}\xspace}
\def\mci {\aunit{mCi}\xspace}

\def\sx    {\ensuremath{\sigma_x}\xspace}    
\def\sy    {\ensuremath{\sigma_y}\xspace}   
\def\sz    {\ensuremath{\sigma_z}\xspace}    

\newcommand{\stat}{\aunit{(stat)}\xspace}
\newcommand{\syst}{\aunit{(syst)}\xspace}
\newcommand{\lumi}{\aunit{(lumi)}\xspace}


\def\order{{\ensuremath{\mathcal{O}}}\xspace}
\newcommand{\chisqndf}{\ensuremath{\chi^2/\mathrm{ndf}}\xspace}
\newcommand{\chisqip}{\ensuremath{\chi^2_{\text{IP}}}\xspace}
\newcommand{\chisqfd}{\ensuremath{\chi^2_{\text{FD}}}\xspace}
\newcommand{\chisqvs}{\ensuremath{\chi^2_{\text{VS}}}\xspace}
\newcommand{\chisqvtx}{\ensuremath{\chi^2_{\text{vtx}}}\xspace}
\newcommand{\chisqvtxndf}{\ensuremath{\chi^2_{\text{vtx}}/\mathrm{ndf}}\xspace}

\def\deriv {\ensuremath{\mathrm{d}}}

\def\gsim{{~\raise.15em\hbox{$>$}\kern-.85em
          \lower.35em\hbox{$\sim$}~}\xspace}
\def\lsim{{~\raise.15em\hbox{$<$}\kern-.85em
          \lower.35em\hbox{$\sim$}~}\xspace}

\newcommand{\abs}[1]{\ensuremath{\left\|#1\right\|}} 
\newcommand{\Real}{\ensuremath{\mathcal{R}e}\xspace}
\newcommand{\Imag}{\ensuremath{\mathcal{I}m}\xspace}

\def\PDF {PDF\xspace}

\def\sPlot{\mbox{\em sPlot}\xspace}
\def\sFit{\mbox{\em sFit}\xspace}


\def\Ebeam {\ensuremath{E_{\mbox{\tiny BEAM}}}\xspace}
\def\sqs   {\ensuremath{\protect\sqrt{s}}\xspace}
\def\sqsnn {\ensuremath{\protect\sqrt{s_{\scriptscriptstyle\text{NN}}}}\xspace}
\def\pt         {\ensuremath{p_{\mathrm{T}}}\xspace}
\def\ptsq       {\ensuremath{p_{\mathrm{T}}^2}\xspace}
\def\ptot       {\ensuremath{p}\xspace}
\def\et         {\ensuremath{E_{\mathrm{T}}}\xspace}
\def\mt         {\ensuremath{M_{\mathrm{T}}}\xspace}
\def\dpp        {\ensuremath{\Delta p/p}\xspace}
\def\msq        {\ensuremath{m^2}\xspace}
\newcommand{\dedx}{\ensuremath{\mathrm{d}\hspace{-0.1em}E/\mathrm{d}x}\xspace}
\def\dllkpi     {\ensuremath{\mathrm{DLL}_{\kaon\pion}}\xspace}
\def\dllppi     {\ensuremath{\mathrm{DLL}_{\proton\pion}}\xspace}
\def\dllepi     {\ensuremath{\mathrm{DLL}_{\electron\pion}}\xspace}
\def\dllmupi    {\ensuremath{\mathrm{DLL}_{\muon\pi}}\xspace}

\def\degrees{\ensuremath{^{\circ}}\xspace}
\def\murad{\ensuremath{\,\upmu\nospaceunit{rad}}\xspace}
\def\mrad{\aunit{mrad}\xspace}
\def\rad{\aunit{rad}\xspace}

\def\betastar {\ensuremath{\beta^*}}
\newcommand{\lum} {\ensuremath{\mathcal{L}}\xspace}
\newcommand{\intlum}[1]{\ensuremath{\int\lum=#1}\xspace}  


\def\bcvegpy    {\mbox{\textsc{Bcvegpy}}\xspace}
\def\boole      {\mbox{\textsc{Boole}}\xspace}
\def\brunel     {\mbox{\textsc{Brunel}}\xspace}
\def\davinci    {\mbox{\textsc{DaVinci}}\xspace}
\def\dirac      {\mbox{\textsc{Dirac}}\xspace}
\def\evtgen     {\mbox{\textsc{EvtGen}}\xspace}
\def\fewz       {\mbox{\textsc{Fewz}}\xspace}
\def\fluka      {\mbox{\textsc{Fluka}}\xspace}
\def\ganga      {\mbox{\textsc{Ganga}}\xspace}
\def\gaudi      {\mbox{\textsc{Gaudi}}\xspace}
\def\gauss      {\mbox{\textsc{Gauss}}\xspace}
\def\geant      {\mbox{\textsc{Geant4}}\xspace}
\def\hepmc      {\mbox{\textsc{HepMC}}\xspace}
\def\herwig     {\mbox{\textsc{Herwig}}\xspace}
\def\moore      {\mbox{\textsc{Moore}}\xspace}
\def\neurobayes {\mbox{\textsc{NeuroBayes}}\xspace}
\def\photos     {\mbox{\textsc{Photos}}\xspace}
\def\powheg     {\mbox{\textsc{Powheg}}\xspace}
\def\pythia     {\mbox{\textsc{Pythia}}\xspace}
\def\resbos     {\mbox{\textsc{ResBos}}\xspace}
\def\roofit     {\mbox{\textsc{RooFit}}\xspace}
\def\root       {\mbox{\textsc{Root}}\xspace}
\def\spice      {\mbox{\textsc{Spice}}\xspace}
\def\tensorflow {\mbox{\textsc{TensorFlow}}\xspace}
\def\urania     {\mbox{\textsc{Urania}}\xspace}

\def\cpp        {\mbox{\textsc{C\raisebox{0.1em}{{\footnotesize{++}}}}}\xspace}
\def\ruby       {\mbox{\textsc{Ruby}}\xspace}
\def\fortran    {\mbox{\textsc{Fortran}}\xspace}
\def\svn        {\mbox{\textsc{svn}}\xspace}
\def\git        {\mbox{\textsc{git}}\xspace}
\def\latex      {\mbox{\LaTeX}\xspace}

\def\kbit          {\aunit{kbit}\xspace}
\def\kbps        {\aunit{kbit/s}\xspace}
\def\kbytes     {\aunit{kB}\xspace}
\def\kbyps      {\aunit{kB/s}\xspace}
\def\mbit          {\aunit{Mbit}\xspace}
\def\mbps        {\aunit{Mbit/s}\xspace}
\def\mbytes     {\aunit{MB}\xspace}
\def\mbyps      {\aunit{MB/s}\xspace}
\def\gbit          {\aunit{Gbit}\xspace}
\def\gbps        {\aunit{Gbit/s}\xspace}
\def\gbytes     {\aunit{GB}\xspace}
\def\gbyps      {\aunit{GB/s}\xspace}
\def\tbit          {\aunit{Tbit}\xspace}
\def\tbps        {\aunit{Tbit/s}\xspace}
\def\tbytes     {\aunit{TB}\xspace}
\def\tbyps      {\aunit{TB/s}\xspace}
\def\dst        {DST\xspace}


\def\nonn {\ensuremath{\mathrm{{ \mathit{n^+}} \mbox{-} on\mbox{-}{ \mathit{n}}}}\xspace}
\def\ponn {\ensuremath{\mathrm{{ \mathit{p^+}} \mbox{-} on\mbox{-}{ \mathit{n}}}}\xspace}
\def\nonp {\ensuremath{\mathrm{{ \mathit{n^+}} \mbox{-} on\mbox{-}{ \mathit{p}}}}\xspace}
\def\cvd  {CVD\xspace}
\def\mwpc {MWPC\xspace}
\def\gem  {GEM\xspace}

\def\tell1  {TELL1\xspace}
\def\ukl1   {UKL1\xspace}
\def\beetle {Beetle\xspace}
\def\otis   {OTIS\xspace}
\def\croc   {CROC\xspace}
\def\carioca {CARIOCA\xspace}
\def\dialog {DIALOG\xspace}
\def\sync   {SYNC\xspace}
\def\cardiac {CARDIAC\xspace}
\def\gol    {GOL\xspace}
\def\vcsel  {VCSEL\xspace}
\def\ttc    {TTC\xspace}
\def\ttcrx  {TTCrx\xspace}
\def\hpd    {HPD\xspace}
\def\pmt    {PMT\xspace}
\def\specs  {SPECS\xspace}
\def\elmb   {ELMB\xspace}
\def\fpga   {FPGA\xspace}
\def\plc    {PLC\xspace}
\def\rasnik {RASNIK\xspace}
\def\elmb   {ELMB\xspace}
\def\can    {CAN\xspace}
\def\lvds   {LVDS\xspace}
\def\ntc    {NTC\xspace}
\def\adc    {ADC\xspace}
\def\led    {LED\xspace}
\def\ccd    {CCD\xspace}
\def\hv     {HV\xspace}
\def\lv     {LV\xspace}
\def\pvss   {PVSS\xspace}
\def\cmos   {CMOS\xspace}
\def\fifo   {FIFO\xspace}
\def\ccpc   {CCPC\xspace}

\def\cfourften     {\ensuremath{\mathrm{ C_4 F_{10}}}\xspace}
\def\cffour        {\ensuremath{\mathrm{ CF_4}}\xspace}
\def\cotwo         {\ensuremath{\mathrm{ CO_2}}\xspace} 
\def\csixffouteen  {\ensuremath{\mathrm{ C_6 F_{14}}}\xspace} 
\def\mgftwo     {\ensuremath{\mathrm{ Mg F_2}}\xspace} 
\def\siotwo     {\ensuremath{\mathrm{ SiO_2}}\xspace} 

\newcommand{\eg}{\mbox{\itshape e.g.}\xspace}
\newcommand{\ie}{\mbox{\itshape i.e.}\xspace}
\newcommand{\etal}{\mbox{\itshape et al.}\xspace}
\newcommand{\etc}{\mbox{\itshape etc.}\xspace}
\newcommand{\cf}{\mbox{\itshape cf.}\xspace}
\newcommand{\ffp}{\mbox{\itshape ff.}\xspace}
\newcommand{\vs}{\mbox{\itshape vs.}\xspace}
\newcommand{\phz}{\phantom{0}}

\newcommand{\lhcborcid}[1]{\href{https://orcid.org/#1}{\hspace*{0.1em}\raisebox{-0.45ex}{\includegraphics[width=1em]{figs/orcidIcon.pdf}}}}

\section{Introduction}

\subsection{Preamble}

The heavy flavour decays prioritised at LHCb occur at rates much higher than the processes analysed at other LHC experiments. A novel full-software trigger has been implemented, the purpose of which is to select collision events more likely to be of interest.
The new trigger was implemented during the detector upgrade that occurred during 2018-2022, and makes selections using event reconstruction information. The resulting complexity presents new challenges when optimising trigger performance. Physics retention must be managed equitably across the entirety of LHCb's physics programme, while also filtering the $O(10,000)$ petabytes of raw data down to the $\sim30$ petabytes available for storage on disk \cite{lhcb2014lhcb}.

In Sections \ref{sec_lhcb} and \ref{sec_algorithms}, the details and challenges of LHCb's trigger will be outlined. Section \ref{sec_bandwidth_division} will describe the method used to solve these challenges at the first selection stage (HLT1). Finally, the resulting quantified improvements in trigger performance will be demonstrated and summarised in Sections \ref{sec_results} and \ref{sec_conclusions}.

\subsection{The LHCb Upgrade Trigger}
\label{sec_lhcb}
LHCb is a single-arm forward spectrometer covering the pseudorapidity range $2 < \eta < 5$,
stationed at interaction point number 8 on the LHC ring. 
It is one of the four largest detectors at the `Large Hadron Collider' (LHC) at CERN. The main goal of this detector is to discover new physics by probing differences between matter and antimatter, and studying decays of heavy-flavour hadrons. For more details on the detector layout, see Refs. \cite{lhcb2014lhcb, Aaij:2859353}. The upgraded LHCb detector aims to accumulate 50 fb$^{-1}$ of data by 2034, including the data recorded prior to the upgrade. This is made possible by running the detector at an instantaneous luminosity around five times higher than the original detector, which collected 9 fb$^{-1}$ between 2010-2018 \cite{bediaga2012framework}.

\begin{figure}[ht]
    \centering
    \includegraphics[width=1.0\linewidth]{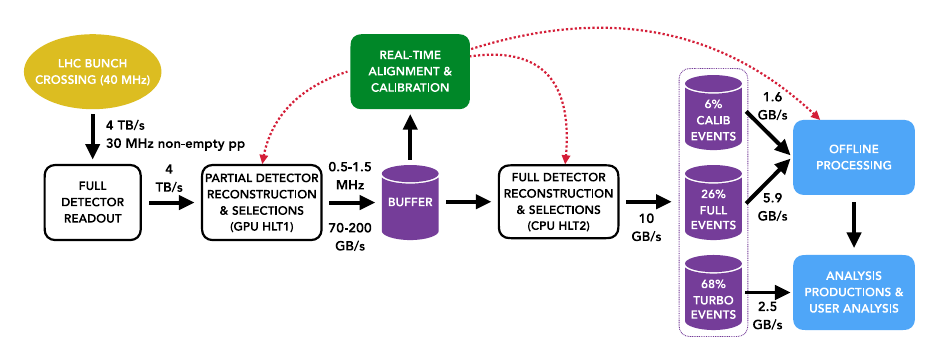}
    \caption[Online-focussed Dataflow]{Online LHCb dataflow \cite{LHCB-FIGURE-2020-016} with ORs obtained from LHCb technical and computing design reports \cite{lhcb2014lhcb, lhcb2018computing}.}
    \label{fig:rta_dataflow}
\end{figure}

The data read out from the detector is managed by Central Processing Unit (CPU) servers that process this information into packets of events that can be processed directly by the High Level Trigger (HLT) at point 8 of the LHC. 
During proton-proton $(pp)$ collisions, the total output of the detector is around 4 TB/s. 
This data is reconstructed and used by the trigger to select signals of interest, after which approximately 10 GB/s is written to permanent offline storage. 
The corresponding data flow is illustrated in Fig.~\ref{fig:rta_dataflow}.

The previous hardware-based first-level trigger made decisions relying on simpler, localised information, such as energy deposits in the calorimeters, prior to readout of the full detector. At larger nominal instantaneous luminosity planned for 2022-2034 data taking, this would saturate. Consequently, the hardware trigger efficiency for hadronic final states dropped as a function of instantaneous luminosity. With the new configuration that makes selections based on track reconstruction, the trigger efficiency scales well \cite{Aaij2020}. More details on the reconstruction are available in Refs.~\cite{lhcb2014lhcb, Aaij:2665946}.

HLT1 runs on $\sim500$ GPUs to reconstruct the triggerless readout from the subdetectors and frontend electronics. The HLT1 trigger menu must be able to efficiently select signals across very wide ranges of rates and energy scales, covering the full breadth of LHCb's physics programme. For nominal running conditions, $\mathcal{O}(100)$ HLT1 selections are performed \cite{Aaij_Vom_Bruch_Evans_2021}. When compared to the 2010-2018 period, it is now much more straightforward for an analyst with a new physics idea to implement a corresponding HLT1 line. This is due to the significant flexibility of the full software trigger \cite{Aaij2020}.
Additionally, GPUs enable the parallelisation of the event loop and parts of the track reconstruction, which makes them ideal for maximising trigger throughput \cite{10634157, Aaij:2859353}.

HLT1 processes data in real time, the output of which is fed into the buffer before it is eventually processed by HLT2. HLT2 performs full event reconstruction on CPUs. It uses full particle identification from the RICH detectors and calorimeter systems to apply selections to fully aligned and calibrated physics objects.

The output of HLT1 must be kept below an upper limit, while maximising the signal efficiency for all physics channels.
In practice, this involves reducing the readout from $30$ MHz to $\sim 1$ MHz \cite{lhcb2014lhcb}.
The reason for this is that HLT2 processes HLT1-filtered data from the buffer days, or sometimes weeks, after the data is processed by HLT1. Consequently, further event reconstruction becomes possible during detector downtime. 

The limit of the buffer is determined by the fact that HLT2 needs to process around half of the output of HLT1 to not fall behind, since protons are collided in the LHC for approximately half of the operational time between shutdowns \cite{lhcb2018computing}. This goal of reducing the HLT1 readout, while simultaneously maximising physics retention, can be achieved by building a tool that automatically tunes an appropriate set of selections.

Software was previously developed to reduce and divide the bandwidth of the hardware trigger between physics goals for the collection of data between 2010-2018 \cite{Aaij_2019}. The aim of this project was to develop software that equitably divides the bandwidth for the upgraded HLT1. 
This must be automated as the trigger selections and collision conditions can change regularly. For each change, the software must re-divide the bandwidth to optimise HLT1 data acquisition under the new conditions. This is the first automated bandwidth division applied to a software trigger at a HEP experiment.

\subsection{HLT1 Algorithms}
\label{sec_algorithms}
The software project for HLT1, Allen, contains parameters that can be tuned to modify the output data rate of the trigger. The HLT1 trigger menu is composed of many trigger lines (decision algorithms).
Most of the tuned HLT1 parameters assert requirements on the transverse momentum ($p_{\text{T}}$), impact parameter (IP) or impact parameter significance ($\chi^2_{\text{IP}}$). $p_{\text{T}}$ is the momentum of a track in the plane transverse to the LHCb detector's
beam axis. IP is the shortest transverse distance between a track and a vertex, typically the proton-proton collision, and $\chi^2_{\text{IP}}$ is this quantity divided by its uncertainty.

\begin{figure}[ht]
	\centering
    \hspace{-0.13cm}\includegraphics[width=0.49\linewidth]{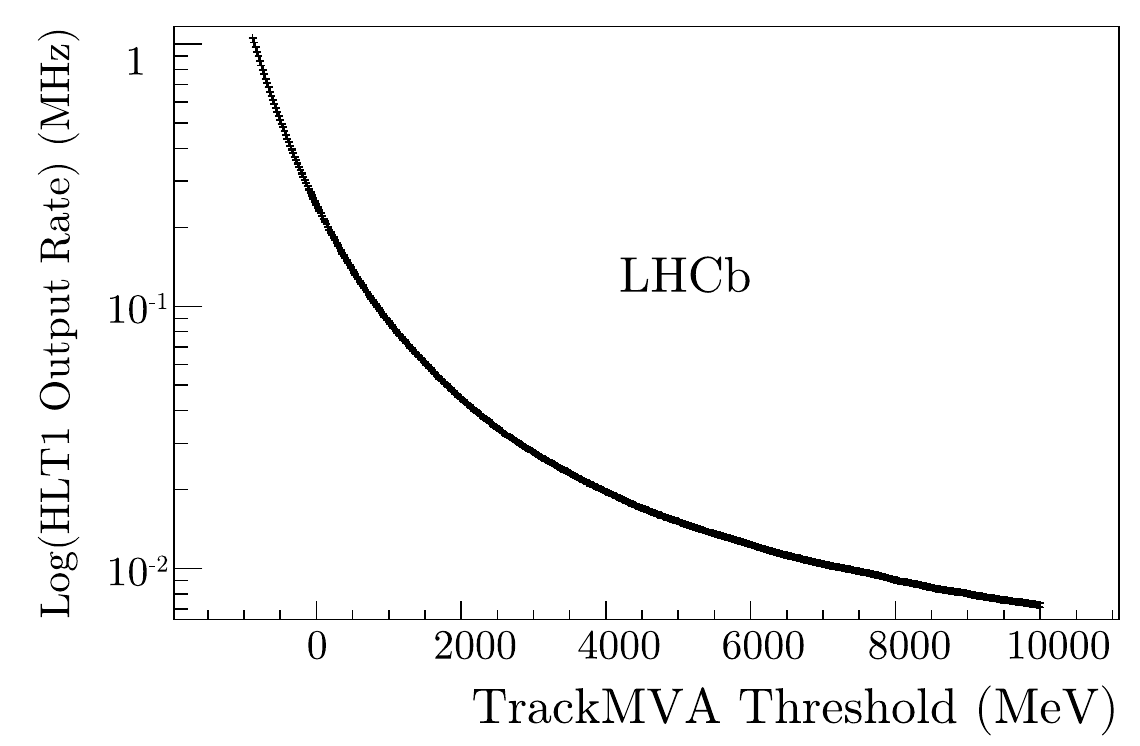}
    \includegraphics[width=0.49\linewidth]{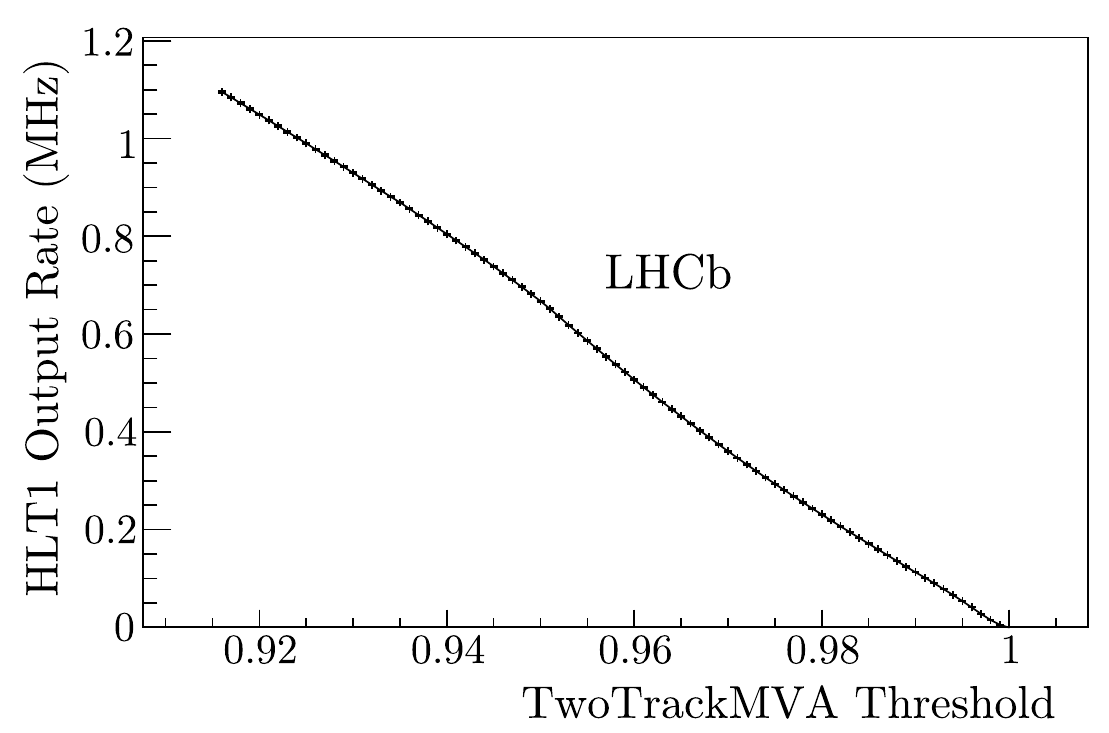}
	\caption[TrackMVA and TwoTrackMVA Inclusive Line Rate vs. Parameter Cut]{\label{trackMVA_parameters}
    HLT1 ORs of the single (left) and two (right) track MVA lines as a function of their input thresholds.
    }
	\label{fig:alphahadrontwotrackmvaratevscut}
\end{figure}

These quantities are useful for distinguishing signal from background because tracks originating from $pp$ collisions typically have small IPs, given by the finite resolution of the vertex locator. 
Conversely, particles originating from heavy-flavour decays will have larger IPs of around $100\mum$ and typically have $p_{\text{T}}$ greater than $2\gevc$. 
Therefore, large numbers of uninteresting events can be rejected by placing requirements on these quantities. 
Cutting too tightly on these quantities will produce an undesirably low efficiency for certain signal channels, and too loosely produces too much OR from HLT1.



The HLT1 trigger lines can be multivariate and can be categorised into inclusive and exclusive selections. 
The majority of the OR comes from the inclusive hadron lines, which select most of the decay channels involving beauty hadrons. 
High signal efficiencies (between 65-95$\%$ at 1 MHz HLT1 output) for beauty decays can be achieved using inclusive selections. 
Conversely, some channels cannot achieve the same quality of physics retention using only the inclusive lines. For example, it is sometimes necessary to use exclusive lines that select a single channel or a mixture of inclusive and exclusive to achieve good efficiencies when selecting certain charm and strange hadrons decays.
Exclusive trigger lines are made possible by the event reconstruction in HLT1. Many unique selections can be performed in real time.

\begin{figure}[ht]
	\centering	\includegraphics[width=0.52\linewidth]{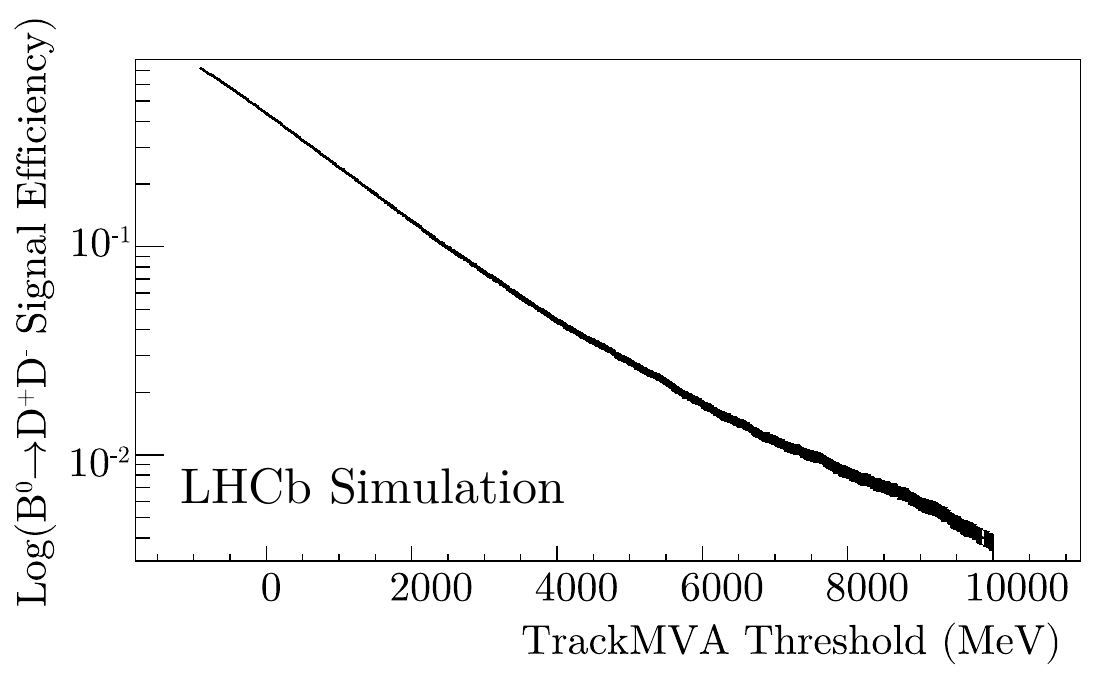}
    \includegraphics[width=0.46\linewidth]{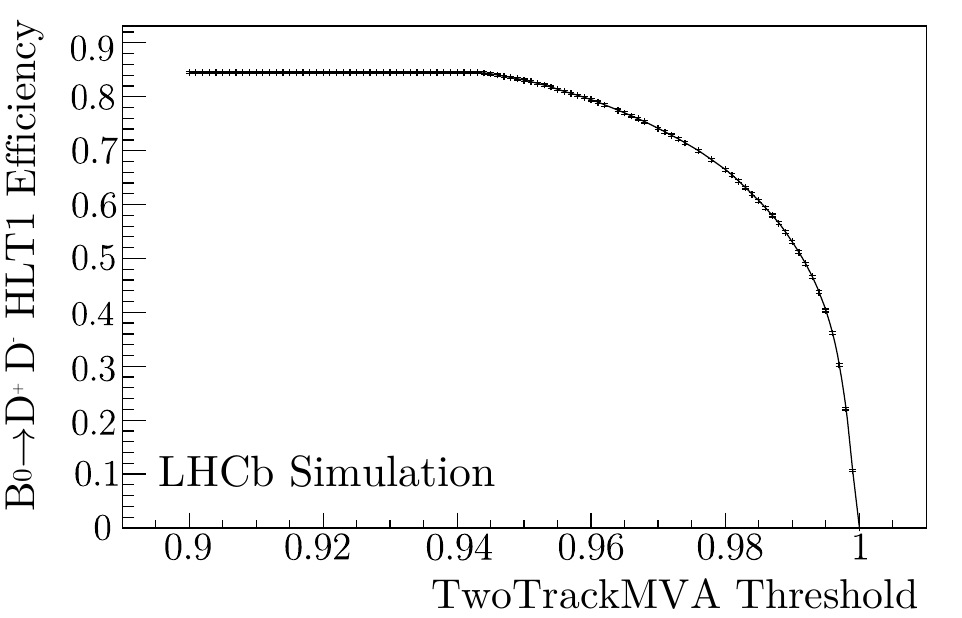}
	\caption{ \label{BtoDD}$B^0 \rightarrow D^+ D^-$ efficiencies of the single (left) and two (right) track MVA lines as a function of their input thresholds.}
	\label{fig:bctodpdmvsalphahadrontwotrackmvacut}
\end{figure}

The inclusive TrackMVA and TwoTrackMVA trigger lines select the largest number of physics signals and are responsible for the majority of the HLT1 OR. For that reason, modifying the value of their input thresholds produces the largest effect on the performance of HLT1. It is important that the OR and efficiency curves of these lines, which are instrumental to any figure-of-merit chosen for trigger optimisation, are smooth. This significantly increases the convergence rate.

The TrackMVA algorithm performs a preselection of tracks to remove the most obvious fakes, and then applies a hyperbolic selection criterion to a two-dimensional plane of $p_{\text{T}}$ and $\chi^2_{\text{IP}}$. This method captures most of the signal background discrimination and is simple to implement. The TwoTrackMVA algorithm is designed to identify pairs of tracks originating from the decays of beauty or charm hadrons using a multivariate classifier \cite{Aaij2020, rta_tdr}.

To visualise part of the parameter space that would be explored during optimisation, examples of the OR for these lines with respect to their input thresholds have been provided in Fig.~\ref{fig:alphahadrontwotrackmvaratevscut}. The OR is calculated using $\sim9$ million events of $pp$ collision data. This corresponds to $\sim0.5$ seconds of 2024 LHCb data acquisition using a minimally biasing trigger, with data quality checks at nominal instantaneous luminosity. No other selections were applied.
Examples of the efficiency to select $B^0 \rightarrow D^+ D^-$ decays with respect to the requirements on the same lines are shown in Fig.~\ref{fig:bctodpdmvsalphahadrontwotrackmvacut}.

This efficiency is calculated from a simulation sample generated from 2024 run conditions, containing simulated events that were selected by at least one representative trigger line and possessed protons in both crossing bunches. Again, no other selections were applied. Each signal sample contains sufficient simulated $pp$ collision events ($\sim100$k) such that the uncertainty on the relative efficiency is much smaller than the change in the efficiency when modifying the values of the discrete line input thresholds. The lines shown in Fig.~\ref{fig:bctodpdmvsalphahadrontwotrackmvacut} are the only ones that select this signal channel. The ORs and efficiencies for Figs.~\ref{fig:adamvsga_scatter}-\ref{fig:withutthresholds1200} are calculated in the same way.

To automatically tune these efficiencies and the HLT1 OR for the purpose of optimising trigger performance, a pseudo-$\chi^2$ figure-of-merit was proposed.


    \section{Method}
\label{sec_bandwidth_division}

\subsection{Figure-of-Merit}

At the beginning of this project, the LHCb Collaboration chose a set of signal channels from which the physics retention of a given trigger selection could be calculated. This ensemble is a subset of the total number of channels being studied at LHCb. Increasing the number of channels selected by a given trigger line or weighting those channels more favourably would lead to this line being loosened further during tuning. Hence, these channels and their corresponding figure-of-merit weightings were chosen carefully to best represent the physics interests of the collaboration. 

A set of sixteen floating HLT1 input thresholds, denoted $\textbf{x}$, were chosen to modify the efficiencies of the representative channels and the HLT1 OR. The HLT1 thresholds excluded from $\textbf{x}$ are either fixed, or they are not applied and their corresponding trigger lines are not included in the bandwidth division.
A small number of thresholds and lines are not optimised due to their negligible contribution to the total rate. A pseudo-$\chi^2$ figure-of-merit is minimised in order to tune $\textbf{x}$. 

The trigger efficiency for each signal channel is measured on a simulated sample of data in which each event contains a reconstructible signal associated with that physics channel. Reconstructibility requirements are applied to the simulated events because most of them are generated without requiring that every track traverses the entire tracking system. This ensures that the bandwidth division determines thresholds based on signal efficiencies for candidates that would be selected at HLT1.
A signal event is considered reconstructible when all of the charged particles from the signal decay have a corresponding reconstructible track. 
A track is reconstructible when it has sufficient hits in the tracking system, on both sides of LHCb's magnet, to be reconstructed \cite{lhcb2014lhcb}. 
The trigger efficiency is then the subset of these events that fire on at least one representative trigger line divided by the number of simulated, reconstructible events with protons in both crossing bunches.

If the OR of the trigger exceeds a certain limit ($\text{OR}_{\text{limit}}$), the trigger efficiencies included in the $\chi^2$ are penalised. The trigger efficiencies are included to account for the loss of physics when constraining the HLT1 OR. The OR is calculated from a sample of `minimally biased' data collected by the detector during 2024, consisting of $9$ million events with at least one proton-proton collision: $$\text{OR} = \frac{N^{\text{passed any}}(\textbf{x})}{N^{\text{total any}}}\times\text{event rate}.$$ The event rate is the number of events per second that HLT1 processes at nominal luminosity during data taking ($\sim30$ MHz). $\text{OR}_{\text{limit}}$ is usually $1$ MHz.
A range of tunings at different rate limits (typically between $0.5$ and $1.5\mhz$) enables the collaboration to adapt to changing physics conditions. 
For example, changes to the nominal luminosity due to gains in throughput from HLT2 would necessitate the ability to interpolate from a range of thresholds.

The events for $N^{\text{passed any}}$ can pass any or all of the HLT1 trigger algorithms included in the division. This means that the $N^{\text{passed any}}(\textbf{x})$ depends on all of the elements of $\textbf{x}$. The $\chi^2$ represents the weighted sum of the loss of physics retention across the chosen set of signal channels and is given by

\begin{equation*}
\label{eq_loss_func}
\chi^2_{\text{global}}(\textbf{x}) = \sum_{i}^{\text{channels}}\omega_i  \left(1 - \frac{\epsilon_i(\textbf{x})}{\epsilon^{\text{max}}_i}\right)^2,
\end{equation*} 
where $\omega_{\text{i}}$ represents the relative importance of each channel, and is usually set to one. 
The rate-penalised efficiency for the $i$th signal decay mode, $\epsilon_i(\textbf{x})$, is
\begin{equation*}
\epsilon_{i}(\textbf{x}) = \begin{cases}
	\frac{N_{i}^{\text{passed}}(\textbf{x})}{N_{i}^{\text{total}}} & \text{OR} \leq \text{OR}_{\text{limit}}, \\
	\frac{N_{i}^{\text{passed}}(\textbf{x})}{N_{i}^{\text{total}}}\times\frac{\text{OR}_{\text{limit}}}{\text{OR}} & \text{otherwise},
\end{cases}
\end{equation*}
where $N_{i}^{\text{passed}}(\textbf{x})$ is the number of reconstructible signal events passing the decisions of the HLT1 trigger selections that would typically be required by an analysis of the given signal mode.
$N_{i}^{\text{total}}$ is the total number of reconstructible signal events where there are protons in both crossing bunches.
$\epsilon_i^{\text{max}}$ is the rate-penalised efficiency for a given channel if the entirety of the bandwidth is allocated to it. 
This is the highest efficiency achievable at HLT1. 
$\epsilon_i^{\text{max}}$ is calculated before the global minimisation using a separate figure-of-merit, but the same rate-penalised efficiencies, and is given by $\chi^2_{\text{indiv}}(\textbf{x}) = (1-\epsilon_{\text{indiv}}(\textbf{x}))^2$.

The chosen set of continuous thresholds $\textbf{x}$ must be discretised to a grid of possible solutions. The step sizes were chosen to be sufficiently large to ensure statistical significance, given the finite number of available events. It was necessary to truncate the continuous thresholds to discrete values to avoid tuning on statistical noise in the data. 

Tuning of the OR and signal efficiencies is currently performed over 35 trigger lines for 80 physics channels chosen by analysts.
Each channel can be selected by several thresholds/trigger lines. Thresholds can be shared across multiple lines to ensure selections are consistent between control and signal modes. 
More than one threshold can also be input into a singular trigger line. A specialised configuration of Allen was used to produce output files that contain the minimal amount of information required to reproduce and modify the settings of the trigger decision.
The event information for each signal sample and the minimally biased sample is read from the Allen output files and stored in C++ objects.

The division of the new full software trigger provides significant computational complexity when compared to the bandwidth division of the hardware trigger during 2010-2018. 
One of the goals of this project is to enable users to quickly run the entire minimisation process and obtain from the bandwidth division tool an optimal set of thresholds using modest computing resources.
The OR and signal efficiencies are recalculated tens of thousands of times. This necessitated optimisations to the following functions which resulted in a run time speedup from several hours per $\chi^2_{\text{indiv}}(\textbf{x})$ minimisation to one to two minutes:

\begin{itemize}
    \item Automatic removal of candidates outside of the desired phase space. If a signal candidate is only accessible by loosening the desired lines to the point where the HLT1 OR is greater than $1.1$ MHz, this candidate is removed.
    \item Reading tabulated ROOT event data into memory (C++ data structures) for faster event filtering during efficiency calculations \cite{root_cite}.
    \item Parallelising the $\chi^2$ evaluations over multiple threads using OpenMP \cite{dagum1998openmp}.
\end{itemize}

\subsection{Choosing a Minimisation Algorithm}
\label{sec_min_alg}

The stochastic `Genetic Algorithm' (GA) \cite{Mitchell1996AnIT} was employed to find the optimal selection when the problem was less CPU-intensive, during the 2010-2018 data taking period. 
This method was chosen due to the fact that the GA operates with input parameters that are constrained to a discrete grid of coordinates. Therefore, the GA required almost no adaptation to minimise the $\chi^2$.

\begin{figure}[ht]
    \centering
    \includegraphics[width=1.0\linewidth]{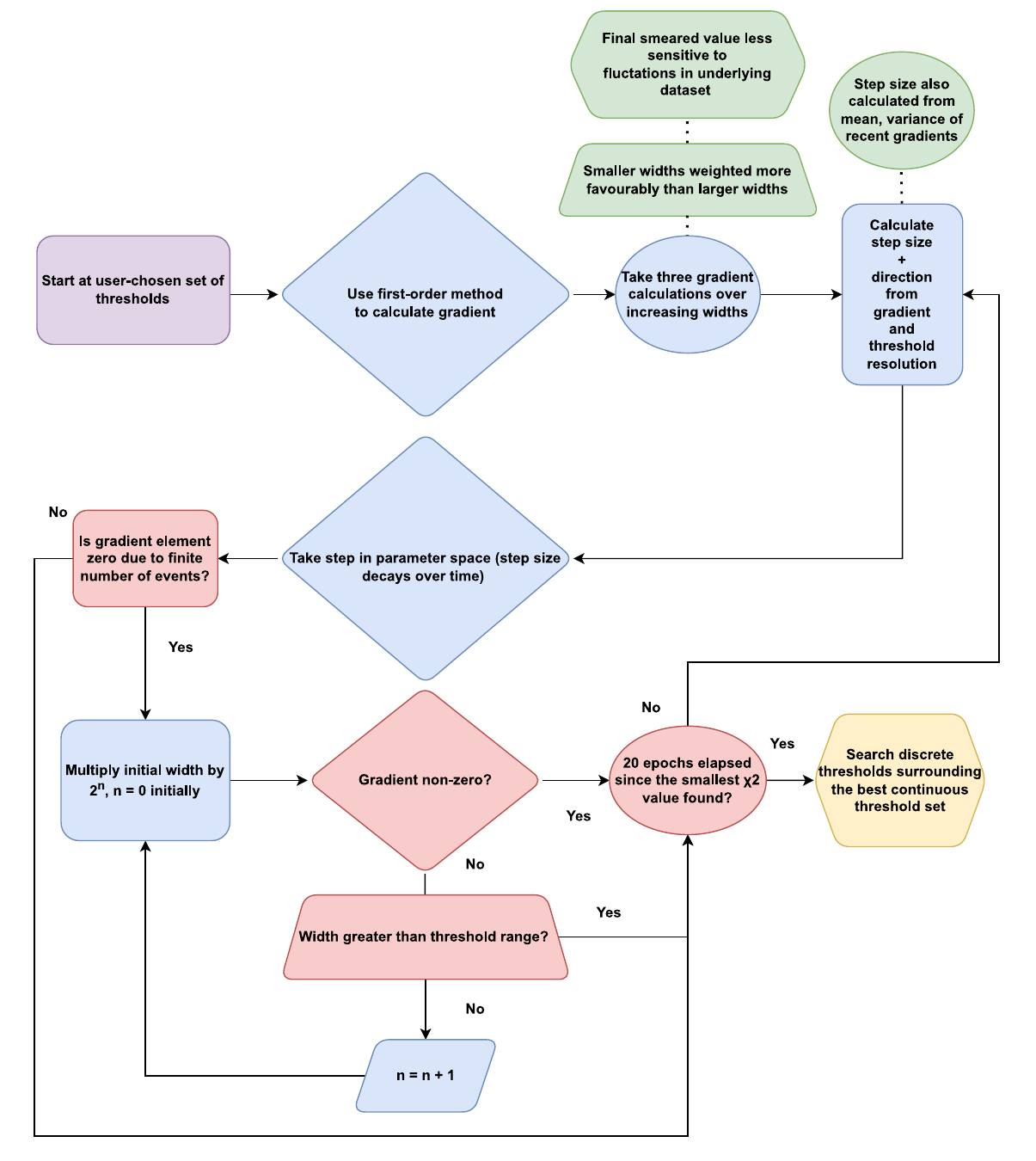}
    \caption{Flow chart of the Adam algorithm adapted to search for the best set of truncated thresholds in a discrete $\chi^2$ space \cite{drawio}.}
    \label{fig:adam_flow_chart}
\end{figure}

However, this requires a careful choice of hyperparameters to balance run time with search depth for a given number of dimensions and possible solutions. 
This is even more important when performing a minimisation for nominal running conditions during 2022-2034. There are approximately four times the number of samples included in the $\chi^2_{\text{global}}$ calculation, and three times the number of parameter dimensions when compared to the version of the bandwidth division tool used during 2011-2018.
Hence, the gradient-based Adapted-Moment (Adam) algorithm \cite{Ketkar2017} has been chosen to improve performance and reduce the number of $\chi^2$ calculations required before finding the global minimum. The workflow of the Adam algorithm is illustrated in Fig.~\ref{fig:adam_flow_chart}.

Adam is computationally efficient and is well-suited to problems with large datasets, many parameters, and noisy gradients. All three of these challenges are encountered when minimising the loss function. Adam is less sensitive to statistical noise in gradient calculations as a first order method than methods that rely on computations of the Hessian, e.g., HESSE from MINUIT \cite{James:2004xla}.

The Adam algorithm is adapted to search for the best continuous selection ensemble before evaluating nearby thresholds constrained to a discrete grid. `Continuous' here means the thresholds can float as any value within a certain range, i.e., not constrained to a discrete grid of possible thresholds.

It was found initially that the Adam algorithm outperformed the GA by at least an order of magnitude in run time to find the best set of thresholds. This comparison was made after choosing appropriate GA hyperparameters, and tuning only five thresholds instead of the final number of $16$.
Adam converged on the solution faster due to the gradient-based search and the fact that searching neighbouring discrete thresholds took less time for less dimensions. 

However, at larger numbers of dimensions and possible solutions, the stochastic GA approach yields similar levels of performance. The bandwidth division tool takes a longer time evaluating the discrete grid of solutions surrounding the continuous solution found by Adam. 
Additionally, the gradient calculations at each epoch become more costly for higher dimensions ($N_{\chi^2} = 6N_{\text{dim}} = 96$ evaluations per epoch).

\begin{figure}
    \centering
    \includegraphics[width=0.75\linewidth]{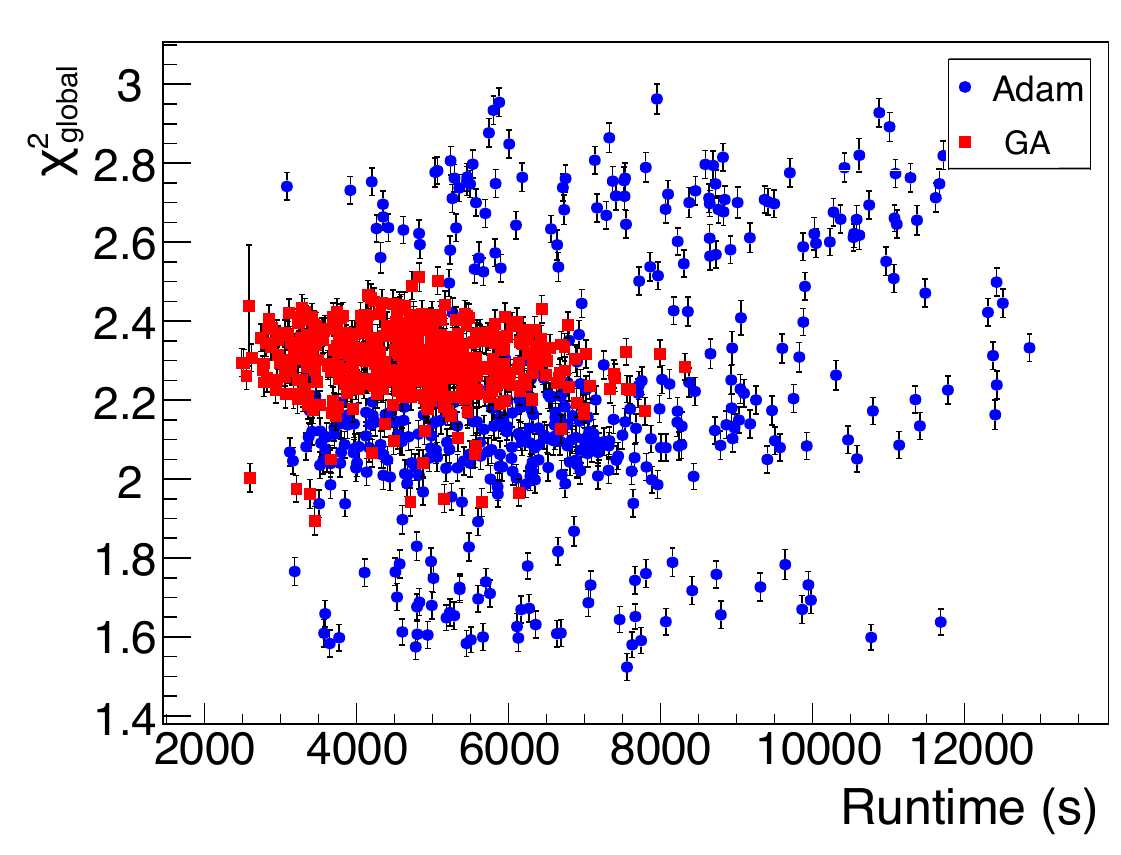}
    \caption[Adam vs. GA Comparison]{A plot of the discrete $\chi^2_{\text{global}}$ vs. algorithm runtime for a number of Adam and GA minimisations starting with different seeds.}
    \label{fig:adamvsga_scatter}
\end{figure}

Fig.~\ref{fig:adamvsga_scatter} compares the performance and speed of the Adam algorithm to the GA over many minimisations. Adam takes longer than the GA, but finds better solutions on average, despite starting from random sets of coordinates each time. Across all samples, there is an average increase in efficiency of $\sim0.4\%$ between the average $\chi^2$ value of the GA and the average $\chi^2$ value of Adam. However, the standard deviation of the GA $\chi^2$ is $\sim70\%$ smaller than that of Adam. The average GA runtime is also $\sim30\%$ shorter than the average Adam runtime.

Adam is preferred due to the need to carefully choose the GA hyperparameters. 
Another advantage of Adam is the fact that the starting thresholds for the Adam minimisation can be chosen to be in the vicinity of the thresholds chosen from previous divisions. 
This increases the likelihood that the selected thresholds are relatively similar to the divisions at slightly looser/tighter HLT1 OR limits. 
Conversely, the GA tends to find thresholds with vastly different values when compared to divisions of looser or tighter OR limits due to its stochastic nature.

\begin{figure}
\centering
\hspace{0.25cm}\includegraphics[width=0.73\linewidth]{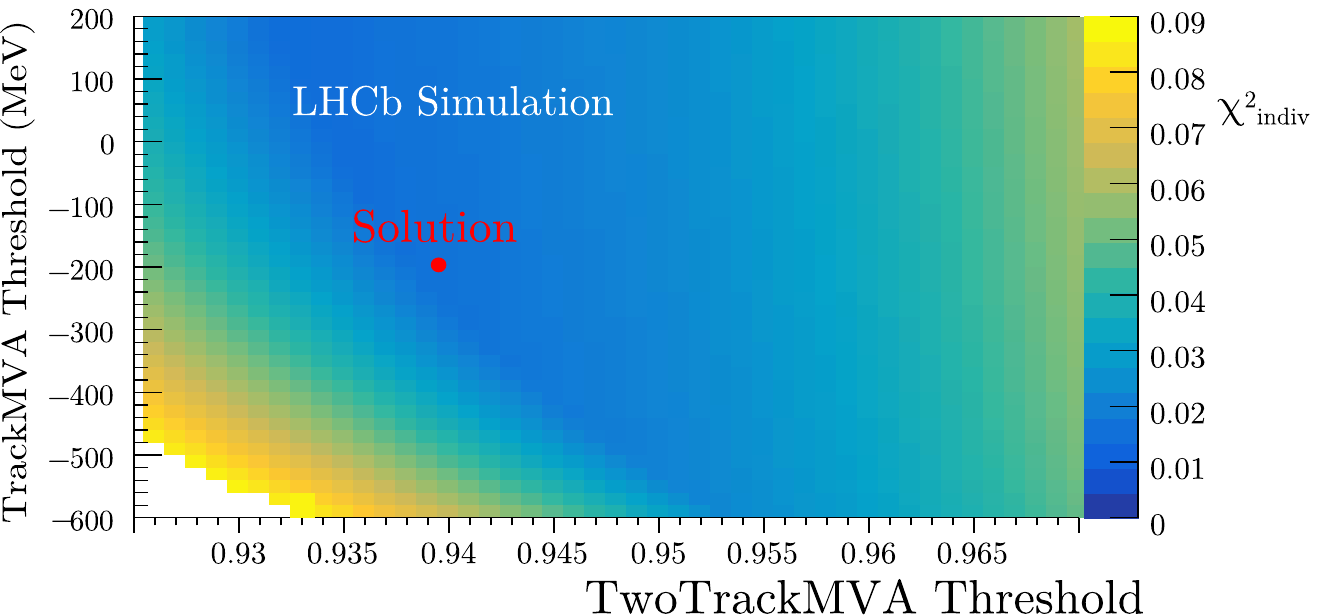}
	\includegraphics[width=0.745\linewidth]{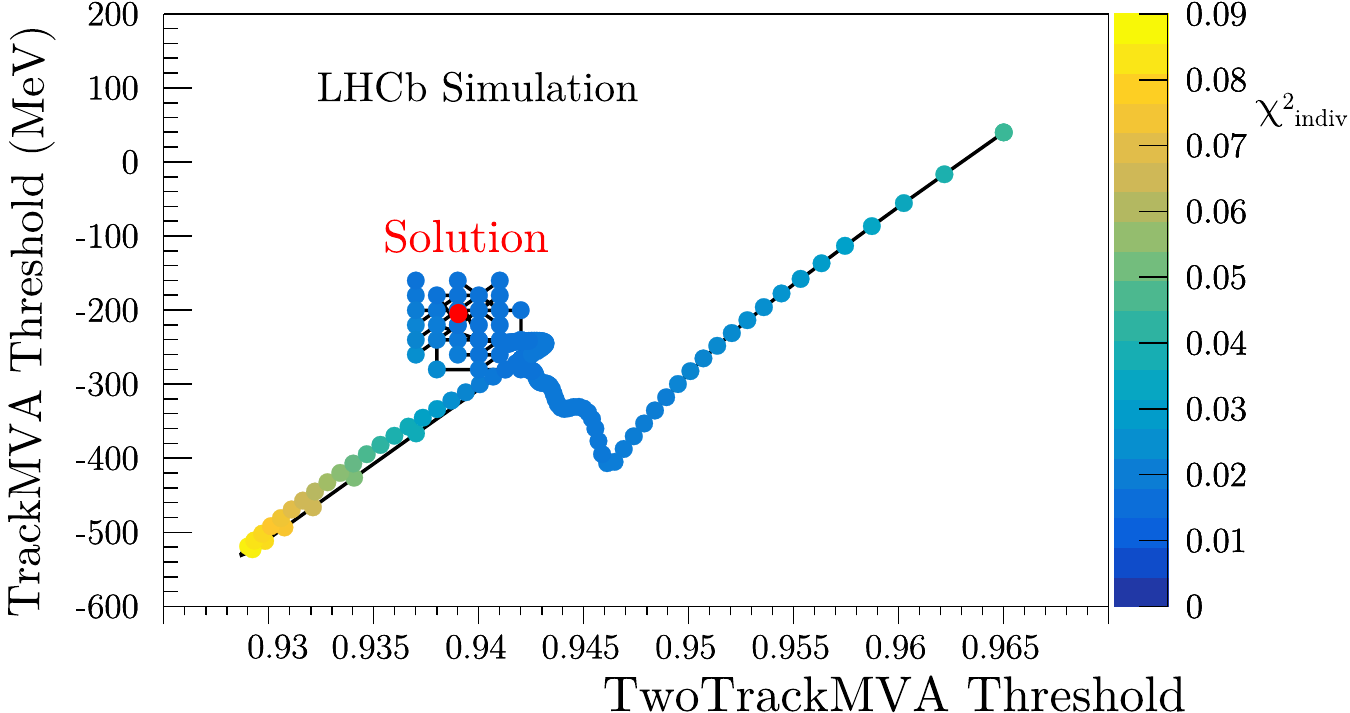}
	\caption[Comparison of $\chi^2_{\text{indiv}}$ Scan and Minimiser Path]{Top - A $\chi^2_{\text{indiv}}$ scan for $B^0 \rightarrow D^+D^-$. Bottom - An analogous minimisation path.}
	\label{fig:fompath_of_b2kstaree}
\end{figure}

An example of the Adam search algorithm in an individual sample minimisation ($\chi^2_{\text{indiv}}$) compared to a scan of the phase space is illustrated in Fig.~\ref{fig:fompath_of_b2kstaree}. Evident in this figure is the discretisation of the continuous solution through a grid search of the nearby discrete thresholds. 

\subsection{Convergence and Discretisation}
\label{choosing_solution_section}

The moment hyperparameters of the Adam algorithm that are responsible for the momentum of the gradient descent are scaled by a heuristic factor of one order of magnitude in the first epoch of the $\chi^2$ minimisation. This helps prevent the minimiser getting stuck in a local minimum if the starting thresholds were situated within one.

\begin{figure}[ht]
	\centering
	\includegraphics[width=0.9\linewidth]{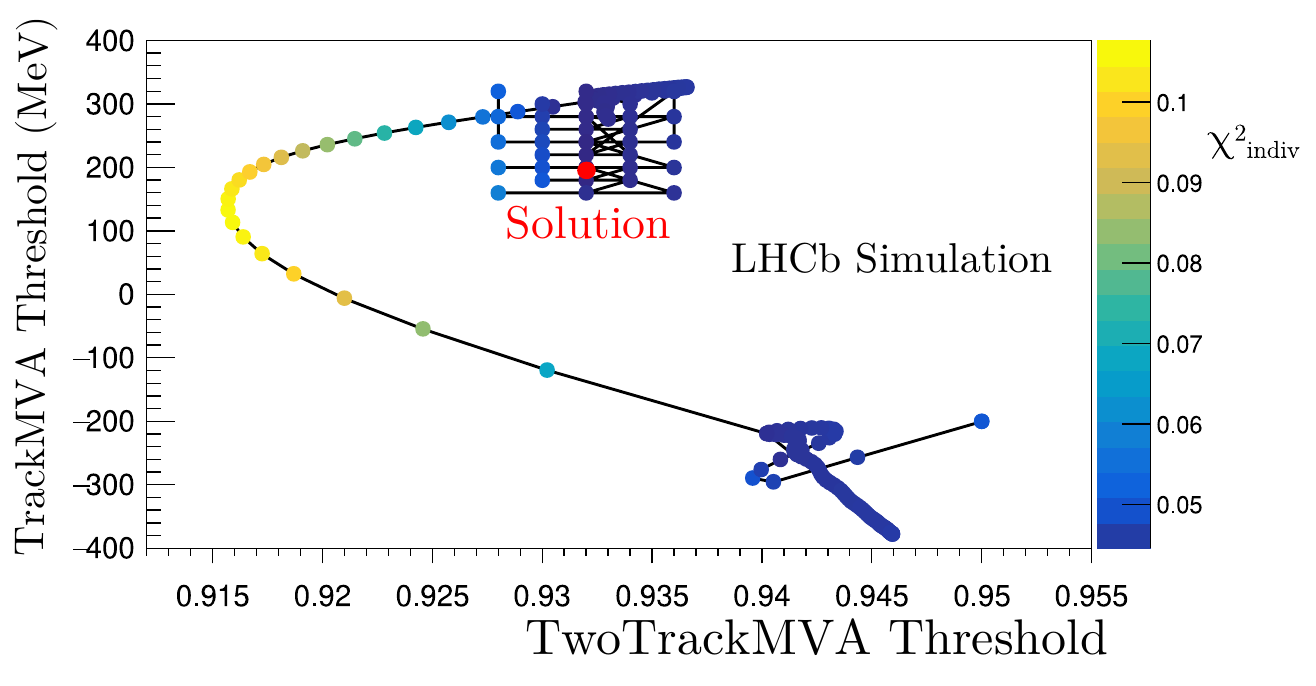}
	\caption[Example of $B^0_{\text{s}}\rightarrow \phi\phi$ Minimisation]{Minimisation of $\chi^2_{\text{indiv}}$ for $B^0_{\text{s}}\rightarrow \phi\phi$.}
	\label{fig:b2kstarmumu_wr}
\end{figure}

After the first Adam minimisation, a `warm-restart' mechanism is initiated. A warm restart involves beginning a second minimisation at the thresholds chosen by the first minimisation with the learning rate for each parameter reset to the starting value. This method was inspired by the minimisers benchmarked in Ref.~\cite{kingma2017adam, loshchilov2019decoupled} and evident in Fig. \ref{fig:b2kstarmumu_wr}. The $\chi^2_{\text{indiv}}$ minimisation path falls into a local minimum and then is `kicked' into a better minimum at a looser TwoTrackMVA threshold and a tighter TrackMVA threshold. Additionally, this figure illustrates the benefit of the Adam algorithm's ability to move against positive gradients during gradient descent.

\begin{figure}[ht]
	\centering
	\includegraphics[width=0.9\linewidth]{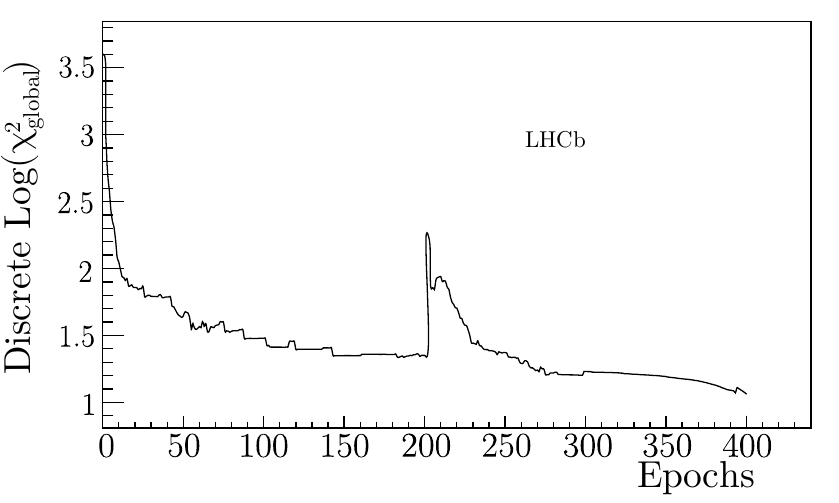}
	\caption[Discrete $log(\chi^2)$ vs. Elapsed Epochs]{A demonstration of the `warm restart' mechanism halfway through the minimisation of $\chi^2_{\text{global}}$.}
	\label{fig:globallogfomepoch}
\end{figure}

Fig.~\ref{fig:globallogfomepoch} also shows the warm restart during the minimisation of $\chi^2_{\text{global}}$. The bandwidth division tool artificially accelerates the search out of a local minimum halfway through the search, resulting in a superior $\chi^2_{\text{global}}$ at the end. The same artificial multiplication of the Adam moments occurs as it does during the first epoch. However, in this epoch the moments are increased in the direction opposite to the previously explored areas of the parameter space. This increases the range of thresholds being searched. 

Once the stopping condition specified in Fig. \ref{fig:adam_flow_chart} is satisfied, the smallest $\chi^2$ value is chosen. The grid thresholds nearest the continuous thresholds would then be located. 
A non-degenerate search of neighbouring discrete selections is performed starting from this nearest set of thresholds to find the smallest discrete $\chi^2$ value. 
The bandwidth division tool would move from threshold set to threshold set, searching neighbouring thresholds to see if a smaller $\chi^2$ value could be found. 
If so, these would be the new central `base' thresholds from which the next group of neighbouring thresholds would be searched. 
The search is performed recursively until a local $\chi^2$ minimum on the grid is obtained.

The first grid search is performed only over the coordinates that can be reached by making one or two `jumps' (translations) to neighbouring sets from the base set. However, jumping twice in the same threshold dimension is prohibited.
This method significantly reduces the run time when compared to evaluating all neighbouring points. However, it also means that the search is more sensitive to the resolution of the grid. 

The continuous thresholds need to be sufficiently close to the grid, i.e., sufficiently granular to avoid an erroneous choice of discrete solution. 
If the grid is too granular then this increases the chance of a worse local minimum being chosen, since the depth of the grid search is reduced.
        \section{Results}
\label{sec_results}
\begin{figure}[ht]
    \centering
    \includegraphics[width=1.0\linewidth]{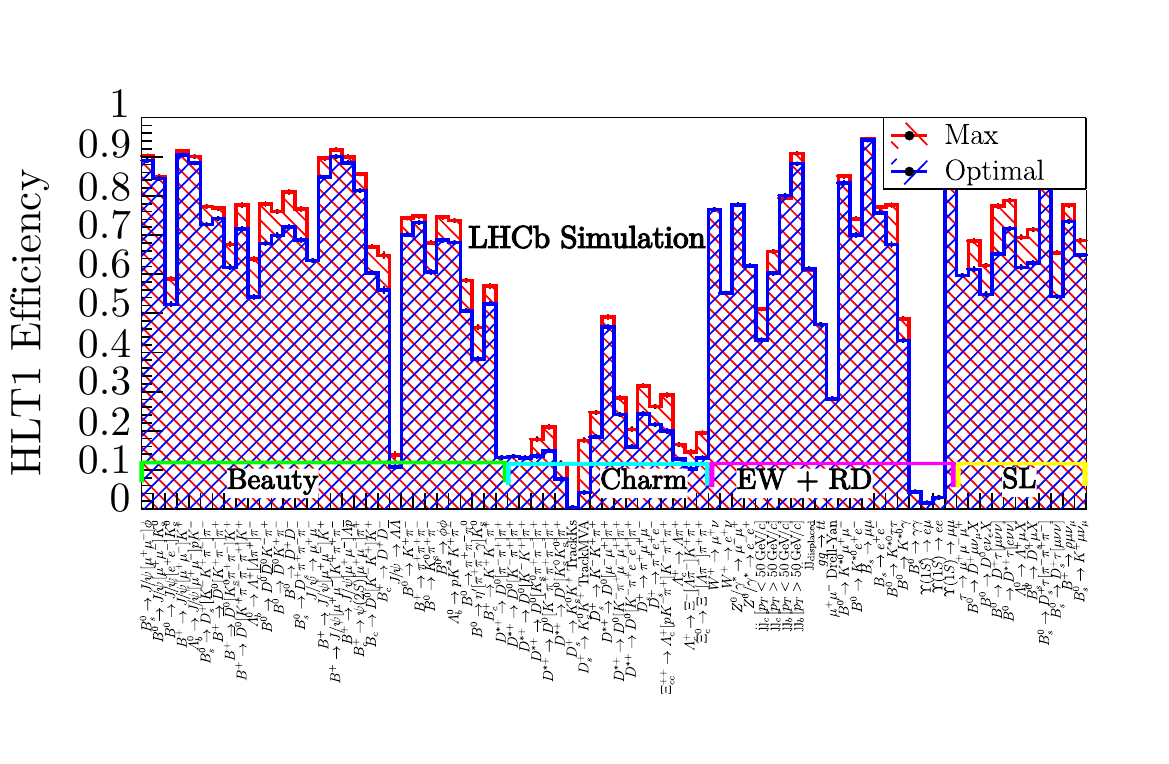}
    \caption[Efficiencies of signal samples using optimal ensemble of thresholds at $1$ MHz]{The optimal (blue) and maximum (red) efficiencies for each signal channel produced by the tool with a $1$ MHz HLT1 output rate limit.}
    \label{fig:withutthresholds1000}
\end{figure}

\begin{figure}[ht]
    \centering
    \includegraphics[width=1.0\linewidth]{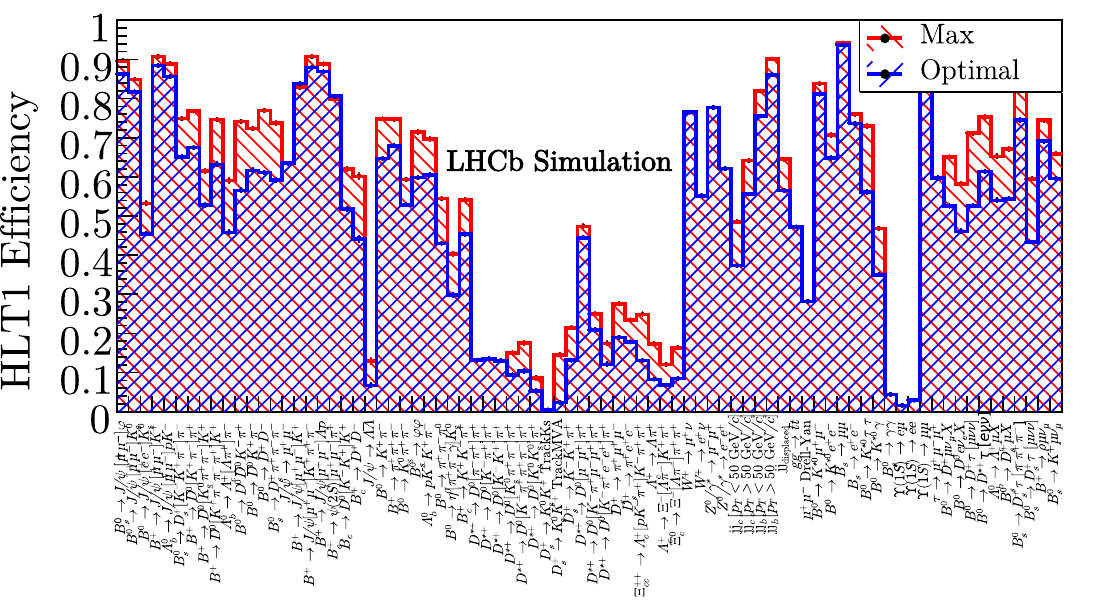}
    \caption[Efficiencies of signal samples using optimal ensemble of thresholds at 800 kHz]{The optimal (blue) and maximum (red) efficiencies for each signal channel produced by the tool at the tightest rate limit (800 kHz).}
    \label{fig:withutthresholds800}
\end{figure}

\begin{figure}[ht]
    \centering
    \includegraphics[width=1.0\linewidth]{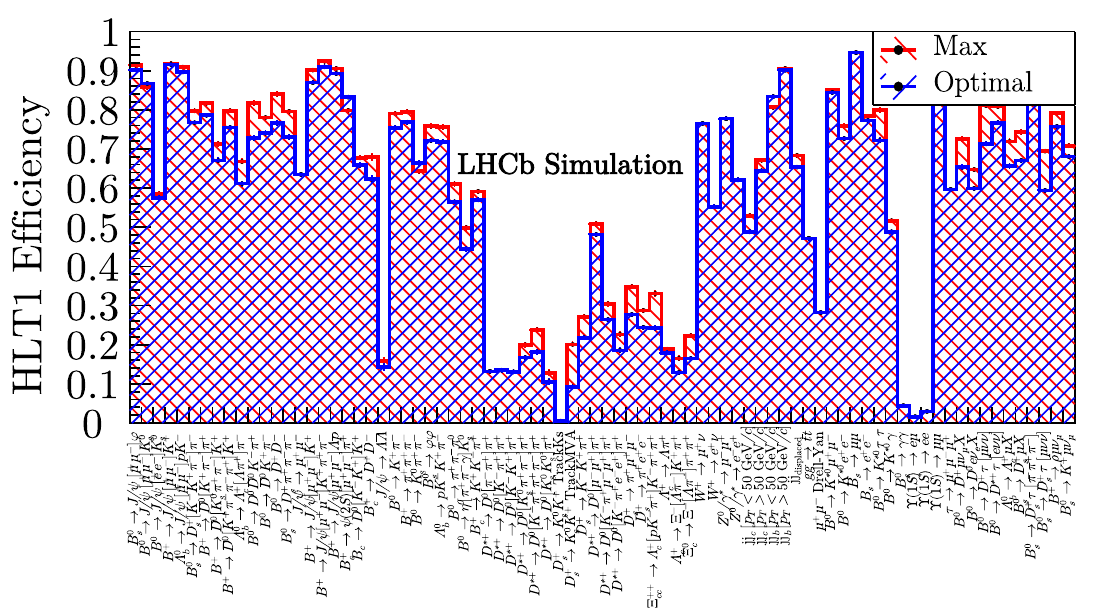}
    \caption[Efficiencies of signal samples using optimal ensemble of thresholds at 1.2 MHz]{The optimal (blue) and maximum (red) efficiencies for each signal channel produced by the tool at the loosest rate limit (1.2 MHz).}
    \label{fig:withutthresholds1200}
\end{figure}

Many bandwidth divisions were performed for various preliminary iterations of the trigger, and also for changes to the physics and operation conditions of the detector. This enabled the collaboration to benchmark HLT1's performance. This occurred before the beginning of the data taking period, and facilitated the study of trigger efficiencies of not just individual signals, but also ensembles of signals studied by different analysis groups. Consequently, refinements were made that led to improvements in trigger performance in 2024.

The results in Fig.~\ref{fig:withutthresholds1000} show the efficiencies of all 80 signal channels when using the thresholds being tuned for a $1~\mhz$ rate limit.  The blue efficiencies were calculated using the optimal ensemble of HLT1 thresholds produced by the tool at a 1 MHz rate limit. The red efficiencies were calculated using the 80 sets of thresholds chosen to maximise the efficiency of each channel. Channels are loosely grouped into those involving beauty, charm, electroweak (EW), rare (RD) and semileptonic (SL) decays.
These thresholds have been used to collect data in HLT1 during 2024.

The results in Figs.~\ref{fig:withutthresholds800} - \ref{fig:withutthresholds1200} show the analogous results for the loosest and tightest (0.8 and 1.2 MHz) rate limits. 
Between these rate limits, there is an average increase in efficiency of around $30\%$. 
The increase in available bandwidth is largely allocated to the inclusive hadronic lines which select the majority of the physics programme.

An average physics retention ratio ($\epsilon_{i}^{\text{optimal}} / \epsilon_{i}^{\text{max}}$) of $93\%$ is achieved for the intermediate $p_{\text{T}}$ beauty channels, since all are selected by inclusive hadronic lines (and some also by the muonic lines). 
The charm channels show a reasonable average retention of $79\%$. Charm decays are high rate and low $p_{\text{T}}$ and so achieve a poorer efficiency for a reasonable share of the bandwidth. 
The electroweak channels demonstrate very efficient signatures, with an average retention of $97\%$. Some are selected by the high $p_{\text{T}}$ leptonic lines and some by the inclusive hadronic lines. 
The semileptonic channels demonstrate a reasonable average retention of $90\%$ when selected by the inclusive hadronic and leptonic TrackMVA lines.

Some lines are not written to maximise efficiency. Alternatively, the channels that are being selected can be particularly difficult to trigger on. The efficiency for such signals increases very slowly as a function of rate. In these cases, the tool naturally favours the lines with larger efficiencies for a reasonable bandwidth and penalises those that do not. 
For example, the decay $B_{\text{s}} \rightarrow \gamma\gamma$ is challenging to trigger as it does not result in any charged tracks and thus can only be triggered by signals from the electromagnetic calorimeter.

The lines selecting the decay $\jpsi\rightarrow\Lz\Lbar$ have poor efficiency because the decay products are not particularly high in $p_{\text{T}}$ and the lines select a high background rate relative to the signal. 
The lines selecting $\Upsilon(1S) \rightarrow \ell \ell$ are fixed-rate and are not tuned by the division.

Default thresholds were chosen manually to be used for data taking before the development of the bandwidth division tool.
The automatically tuned thresholds resulted in an average increase in signal efficiency of $\sim30\%$ for beauty and  $\sim70\%$ for charm and semileptonic channels, at a saving of 200 kHz in OR, with respect to the efficiencies and OR obtained with the default thresholds.

        \section{Conclusions}
\label{sec_conclusions}

LHCb has pioneered readout and event reconstruction on a fully software-based trigger. This trigger reduces the event rate by a factor 30 before full event reconstruction at a level comparable to offline quality. 
The trigger system has changed considerably since the 2015-2018 data taking period.
However, one thing that remains constant is the requirement for the OR of HLT1 to be reduced to a size that HLT2 can handle, whilst providing an equitable division of physics data between analysis groups. 

Adaptations of stochastic and gradient-based methods have been applied to a high-dimensional phase space challenge produced by the full-software trigger of HLT1. The new bandwidth division tool enables the collaboration to adapt to this challenge quickly. The results of this optimisation serve as a proof of concept that could be well suited to other domains.

The bandwidth division tool is designed to accommodate all of the new requirements and features of the upgraded LHCb trigger. It adapts to the increased demand for accuracy and performance. 
This facilitates a fast turnaround when HLT1 needs re-optimising.  
This can be done under different run conditions and a different ensemble of samples, lines and parameters.

The thresholds chosen by the bandwidth division tool during 2024 have acted to equitably minimise the overall difference between the best possible signal efficiencies and the optimised signal efficiencies. This was achieved for a range of HLT1 OR working points.
The main consequence of the bandwidth division was a significant improvement in the sensitivity of measurements performed by analysts on data collected in the next two data-taking periods. 

For future upgrades, the complexity of the trigger menu will increase and the tool will scale accordingly, as it has between the 2010-2018 and 2022-2034 periods.

\section*{Acknowledgements}

The authors acknowledge the work of Miriam Gandelman on an implementation of the bandwidth division developed for the hardware trigger to take data between 2010-2012. We would also like to thank LHCb’s Real-Time Analysis Project for its support, insightful discussions, and for reviewing an early draft of this paper. The authors are grateful to the LHCb computing and simulation teams for producing the simulated LHCb samples used to calculate the physics retention of the signals of interest. 

They extend thanks to the LHCb Physics Planning Group and WG analysts for their expertise in selecting the physics signals and trigger lines to include in the bandwidth division. Finally, the authors show appreciation to their LHCb colleagues, particularly the RTA piquets and software maintainers, for the development and maintenance of LHCb’s nightly testing and benchmarking infrastructure. The authors have been supported by ERC-StG-2019 Beauty2Charm, and acknowledge support from STFC and UKRI.
	
		\bibliographystyle{ieeetr}
	\bibliography{main.bib}
\end{document}